\documentclass[aps,nofootinbib,prd,eqsecnum,10pt]{revtex4-2}
\usepackage{caption,subcaption}
\usepackage{graphicx,xcolor} 
\usepackage{amsmath,amssymb}
\usepackage{braket,slashed}
\usepackage[pdfstartview=XYZ,
bookmarks=true,
colorlinks=true,
linkcolor=blue,
urlcolor=blue,
citecolor=blue,
bookmarks=true,
linktocpage=true, 
hyperindex=true]{hyperref}
\usepackage{orcidlink}
\usepackage[normalem]{ulem}
\newcommand{\La}{\mathcal{L}}
\newcommand{\nn}{\nonumber}
\newcommand{\Ha}{\mathcal{H}}
\newcommand{\p}{^{\prime}}

\newcommand{\tL}{\tilde{\Lambda}}
\newcommand{\lT}{\lambda_{\operatorname{tot}}}

\newcommand{\Real}{\operatorname{Re}}
\newcommand{\Imag}{\operatorname{Im}}

\begin{document}
\title{Effects of spacetime geometry on neutrino oscillation inside a Core-Collapse Supernova}
\author{Indrajit Ghose \orcidlink{0000-0002-8561-4954}}
\email{ghose.meghnad@gmail.com}
\author{Amitabha Lahiri \orcidlink{0000-0001-8113-6345}}
\email{amitabha@bose.res.in}
\affiliation{S.N. Bose National Centre for Basic Sciences, Saltlake, WB, India}

\begin{abstract}
Neutrinos are excellent probes of the inner structures of supernovas. However, an understanding of their dynamics remains incomplete, which is crucial for properly interpreting the detector data. The huge matter density inside a core collapse supernova affects the self-coupling of the neutrinos through a geometrical four-fermion interaction induced by spacetime torsion generated by the fermions themselves. For normal mass hierarchy, the effect is negligible; for inverted mass hierarchy however, we find that the gravitational coupling can significantly alter the flavor dynamics. The possibility of constraining the said interactions through the relative abundance of different flavors of the neutrinos is discussed.

\end{abstract}

\maketitle

\section{Introduction}
The core-collapse supernova (CCSN) is among one of the most intensely studied objects in astrophysics and astroparticle physics~\cite{Colgate:1966ax, Herant:1994dd, Janka:2006fh, Chakraborty:2011gd, Janka:2012wk, Janka:2012sb, Volpe:2023met, Johns:2025mlm}. These supernovae hold the key to the formation of heavier elements ($r$-process)~\cite{Qian:1993dg, Chakraborty:2009ej, Duan:2010af, Martinez-Pinedo:2013jna, Wu:2013gxa, Wu:2014kaa, Sasaki:2017jry, Xiong:2019nvw, Ko:2019xxm, Xiong:2020ntn, George:2020veu}.
{It is usually said that for a successful explosion of a supernova, a delayed shock {revival} mechanism is required. The cooler ultra-dense core of the CCSN, known as the proto-neutron star (PNS) acts as a neutrino bulb.  These emitted neutrinos and antineutrinos revive the stalled supernova explosion~\cite{MEZZACAPPA1999281, Janka:2025tvf, Betranhandy:2022bvr, couch2017mechanism, yamada2024physical}. { We note that the idea of a stalled CCSN explosion has its detractors -- see~\cite{ Soker:2022qxu, Kuroda:2023zbz} for discussion of an alternative viewpoint. However, irrespective of the details of the explosion mechanism, a robust explanation involves understanding the interaction of the outgoing neutrinos with the infalling matter.  } Energy exchange between the neutrinos and antineutrinos and the matter falling towards the PNS depends on the emitted flux's flavor content~\cite{Dasgupta:2011jf}. Neutrino signals from CCSN and their flavor content are an excellent probe for studying these objects, as their feeble interaction with matter lets us look deeper into the center of the CCSN.}


In a CCSN or early universe, (anti)neutrinos comprise a significant portion of the background. In that case, that matter effect on neutrino oscillation will include the matter potential due to oscillating neutrinos themselves~\cite{PANTALEONE1992128}. Due to neutrinos' self-interaction, the oscillation equations become nonlinear~\cite{Pantaleone:prd, KOSTELECKY1993127, Samuel:1993uw, PhysRevD.52.621, Qian:1994wh, Pantaleone:1994ns, Kostelecky:1994dt, PhysRevD.74.123004, PhysRevD.74.105010}. These analyses show some new features of neutrino oscillation in CCSN. For example, self-interaction couples all the neutrinos and anti-neutrinos and we get a ``collective" oscillation which is not seen in terrestrial situations~\cite{Qian:1993dg, Pastor:2001iu, Duan:2010bg, Pehlivan:2011hp, Patwardhan:2022mxg, Hall:2021rbv, DelfanAzari:2019tez, Gava:2008rp, Duan:2010bf}. This allows a small mixing angle to convert the flavor content completely. Once multi-angle propagation of neutrinos is allowed, the self-interaction shows new features like fast pairwise conversion and flavor instability.

The treatment of neutrino oscillation in CCSN has mostly neglected the effects of spacetime geometry in this explosive astrophysical phenomenon, apart from investigations into the decoherence of neutrino wave packets caused by strong gravitational fields~\cite{Chatelain:2019nkf}. 
It was recently proposed in~\cite{Chakrabarty:2019cau} that spacetime can modify neutrino oscillation parameters {through the non-Riemannian part of the connection}. In the terrestrial experiments dealing with neutrino oscillation~\cite{Barick:2023qjq, Barick:2023wxx, Panda:2024qsh}, the matter background has no significant presence of neutrinos. In the core of the supernova, the self-interactions of the neutrinos have to be taken into account for a full quantum mechanical treatment of the neutrino dynamics. {In this work we show that the spacetime geometry contributes to the self-interaction of neutrinos. We will explore the effects of this new contribution to the matter potential and the self-interaction in the oscillation of neutrino flavors.}

The article is organized as follows. We give a brief introduction to Cartan's extension of GR in Sec.~\ref{sec:ferm_curved}. {We show how to accommodate a chiral interaction between fermions and torsion.} This gives us a natural way to include the effect of torsion in the flavor evolution equation. Then in Sec.~\ref{sec:Hamiltonian} we calculate the contribution of the aforementioned interaction in the Hamiltonian and derive the Liouville-von Neumann equations that dictate the flavor dynamics of neutrinos and antineutrinos. Finally, we plot the flavor evolution in Sec.~\ref{sec:uni_den}. We will {see flavor dynamics in the case of a realistic CCSN in Sec.~\ref{sec:real_sn}}. Finally, we conclude in 
Sec.~\ref{sec:conclusion}.

\section{Fermions in curved spacetime}
\label{sec:ferm_curved}

To connect the microphysical descriptions of strong and electroweak force to the macrophysical gravity, we need to embed the quantum mechanical fields which are different representations of the Poincar\'e group on a $4$-dimensional Riemann-Cartan space-time~\cite{Cartan:1923zea, Hehl:1976kj}. {To write a diffeomorphism invariant theory of spinors it is convenient to introduce the spin connection with components  $A_{\mu}{}^{ab}$. The spin connection can be split into two parts as}
\begin{align}
    A_{\mu}{}^{ab}=\omega_{\mu}{}^{ab}+K_{\mu}{}^{ab}\,.
    \label{eq:spin-connection.split}
\end{align}
The part $\omega_{\mu}{}^{ab}$ is the torsion-free Levi-Civita spin connection and the non-Riemannian part denoted by $K_{\mu}{}^{ab}$ is known as the contortion \cite{Barick:2023wxx}. Every curved spacetime is locally flat and at every point in the curved spacetime, we can define four orthonormal axes. The vector fields corresponding to these are $e^{\mu}_{a}$, called tetrads~\cite{Carroll:2004st}. We will use Greek indices to refer to the spacetime and Latin indices to refer to the locally flat tangent space. The metric $g_{\mu\nu}$ of the curved spacetime and the metric $\eta_{ab}$ for vectors in the tangent space are related 
at each point of spacetime by 
\begin{align}
e^{\mu}_{a}e^{\nu}_{b}g_{\mu\nu}=\eta_{ab}
.\end{align}

The inverse of tetrads $e^\mu_a$ are cotetrads $e^{a}_{\mu}$\,, related by
\begin{align}
    e^{\mu}_{a}e^{a}_{\nu}=\delta^{\mu}_{\nu}\,, \qquad\qquad  e^{a}_{\mu}e^{\mu}_{b}=\delta^{a}_{b}\,,
\end{align}
where spacetime indices are lowered and raised by $g_{\mu\nu}$ and its inverse $g^{\mu\nu}$\,, while internal flat space indices are lowered and raised by $\eta_{ab}$ and its inverse $\eta^{ab}$\,.

The Dirac $\gamma$ matrices are constant in flat space --- we will write them as $\gamma^a$ etc. These can be projected
to the background curved spacetime via
\begin{align}
    \gamma^{\mu}=e^{\mu}_{a}\gamma^{a}\,,
\end{align}
{albeit at the cost of making the Dirac gamma matrices spacetime dependent. The local gamma matrices $\gamma^\mu(x)$ follow
the natural extension of the well-known anticommutation relation of the flat space gamma matrices and are given by $[\gamma_{\mu},\gamma_{\nu}]_{+}=2g_{\mu\nu}$.}

Using the tetrads and the spin connection, we can write the  Dirac Lagrangian for fermions minimally coupled to the background spacetime as
\begin{align}
    \La_{\psi}=\frac{i}{2}	\biggl(\bar{\psi}\gamma^\mu\partial_\mu\psi - 
	\partial_\mu\bar{\psi}\gamma^\mu\psi - \frac{i}{4} A_{\mu}{}^{ab} \,
	\bar{\psi}[\sigma_{ab}, \gamma_c ]_{_+} \psi\, e^{\mu c} \biggr)-m\bar\psi\psi\,.
 \label{curved.fermion}
\end{align}
The Ricci scalar can be written in terms of the spin connection $A_{\mu}{}^{ab}$ and the tetrads $e^\mu_a$ as 
\begin{equation}
    R = F_{\mu\nu}{}^{ab} e^\mu_a e^\nu_b\,,   \label{Ricci} 
\end{equation}
where the field strength$F$ {is defined as} 
\begin{equation}
    F_{\mu\nu}{}^{ab} = \partial_\mu A_\nu{}^{ab} - \partial_\nu A_\mu{}^{ab} + A_{\mu}{}^{a}{}_{c} A_{\nu}{}^{cb} -  A_{\nu}{}^{a}{}_{c} A_{\mu}{}^{cb}\,.
\end{equation}
Using Eq.~(\ref{eq:spin-connection.split}), we can then write the action as 
\begin{align}
    S =&  \frac{1}{2\kappa}\int |e| d^4x\, \left(\hat{R} +  e^\mu_a e^\nu_b \partial_{[\mu}K_{\nu]}{}^{ab} 
	+	 e^\mu_a e^\nu_b\left[\omega_{[\mu}, K_{\nu]}{}^{ab}\right]_{-}\right) \notag \\
	 & \qquad  + \int |e| d^4x\, 
	\left(	\frac{1}{2\kappa}  e^\mu_a e^\nu_b\left[K_{[\mu }, K_{\nu]}{}^{ab}\right]_{-}  
	+	\La_\psi \right)\,,
\label{action}
\end{align}
Where $\kappa=8\pi G_N$ and $\hat{R}$ is the Ricci scalar as calculated for the torsionless part of the connection.
The $K$ terms in the first line of this action combine to make a total derivative. 
Then $K$ has a purely algebraic equation of motion, which gives 
\begin{align}
    K_{\mu}{}^{ab}=\frac{\kappa}{8}e^{c}_{\mu}\bar{\psi}[\gamma_{c},\sigma^{ab}]_{+}\psi\,.
\end{align}
This arises from the coupling between the spin connection and the fermions in Eq.~(\ref{curved.fermion}).
But the torsionless connection components are completely known in terms of metric derivatives, which is equivalent to saying that $\omega_\mu{}^{ab}$ is completely known in terms of derivatives of the tetrads and their inverses. 
Thus $K$ is an independent field which may couple to fermions differently from $\omega_\mu{}^{ab}$. 
In particular, $K$ may couple to different species of fermions with different strengths. It may also couple to 
different chiralities independently.
Thus the generic form of the fermionic part of the Lagrangian is~\cite{Chakrabarty:2019cau}
\begin{align}
    \La_\psi = \sum\limits_{i}	&\biggl(\frac{i}{2}\bar{\psi}_i\gamma^\mu\partial_\mu\psi_i - 
	\frac{i}{2}\partial_\mu\bar{\psi}_i\gamma^\mu\psi_i 	+ \frac{1}{8} \omega_{\mu}{}^{ab} e^{\mu c}  \,
	\bar{\psi}_i[\sigma_{ab}, \gamma_c ]_{_+} \psi_i\, - m\bar{\psi_i}\psi_i\,
 \notag \\ &
	+ \frac{1}{8} K_{\mu}{}^{ab} e^{\mu c}
	\left(\lambda^i_{L}\bar{\psi}_{iL} \left[\gamma_c, \sigma_{ab}\right]_{+}\psi_{iL} 
 + \lambda^i_{R}\bar{\psi}_{iR} \left[\gamma_c, \sigma_{ab}\right]_{+}\psi_{iR}\right) \biggr)\,,
	\label{L_psi_all}
,\end{align}
where $i$ runs over all species of fermions.
The solution of the non-Riemannian degrees of freedom coupled to the matter Lagrangian Eq.~\eqref{L_psi_all} is
\begin{equation}\label{chiral.torsion}
	K_{\mu}{}^{ab} = \frac{\kappa}{4}\epsilon^{abcd}e_{c\mu} \sum\limits_i \left(-\lambda^i_{L}\bar{\psi}_{iL}\gamma_d \psi_{iL} + \lambda^i_{R}\bar{\psi}_{iR}\gamma_d \psi_{iR}\right)\,.
\end{equation}
Replacing the contortion in Eq.~\eqref{L_psi_all} by its solution in Eq.~\eqref{chiral.torsion}, we get the Lagrangian to represent the spin-torsion interaction of fermions as
\begin{align}
    \La_\psi = \sum\limits_{i}	&\biggl(\frac{i}{2}\bar{\psi}_i\gamma^\mu\partial_\mu\psi_i - 
	\frac{i}{2}\partial_\mu\bar{\psi}_i\gamma^\mu\psi_i 	+ \frac{1}{8} \omega_{\mu}{}^{ab} e^{\mu c}  \,
	\bar{\psi}_i[\sigma_{ab}, \gamma_c ]_{_+} \psi_i\, - m\bar{\psi_i}\psi_i\biggr)\,
 \notag \\ &
	- \frac{1}{\sqrt{2}}
	\biggl(\sum\limits_{i}\left(-\lambda^i_{L}\bar{\psi}_{iL}\gamma_d \psi_{iL} + \lambda^i_{R}\bar{\psi}_{iR}\gamma_d \psi_{iR}\right)\biggr)^2\, \label{eq:spin-torsion_interaction}
.\end{align}

In the last step, we have absorbed a factor of $\sqrt{\dfrac{3\kappa}{8\sqrt{2}}}$ in the $\lambda$'s. In this framework, the torsion degrees of freedom ceases to exist outside of the matter distribution i.e. they are non-dynamical. However, spin still manifests itself through the modified Einstein equation~\cite{Barick:2023wxx}
\begin{align}
   \hat{G}_{\mu\nu}=\kappa\hat{T}_{\mu\nu}-\frac{\kappa}{\sqrt{2}}g_{\mu\nu}\biggl(\sum\limits_{i}\left(-\lambda^i_{L}\bar{\psi}_{iL}\gamma_d \psi_{iL} + \lambda^i_{R}\bar{\psi}_{iR}\gamma_d \psi_{iR}\right)\biggr)^2 \,\label{eq:modified_efe}.
\end{align}
We have put a hat to indicate that the corresponding object is constructed with a torsion-free connection. The torsion has been entirely integrated out and we are left with a torsion-free Einstein field equation with a modified energy-momentum tensor. 

Just like Eq.~\eqref{eq:modified_efe} gives us the modified Einstein-like Field Equations, Eq.~\eqref{eq:spin-torsion_interaction} gives us the Lagrangian density of the effective torsionless field theory, with an effective four-fermion interaction. If we assume that the neutrinos only couple through the left-handed current in the new quartic interaction, then the part of the last term in Eq. \eqref{eq:spin-torsion_interaction} that contributes to the neutrino self-interaction simplifies to
\begin{align}
\La_{\operatorname{self}}= {-\frac{1}{4\sqrt{2}}\sum_{i,j}\left(\lambda_i \bar{\nu}_i\gamma^{\mu}(1-\gamma_5)\nu_i\right) \left(\lambda_j\bar{\nu}_j \gamma_{\mu}(1-\gamma_5) \nu_j\right)}\label{eq:spin-torsion_self-interaction}
.\end{align}

\section{The Hamiltonian of Neutrinos in a neutrino-rich background}
\label{sec:Hamiltonian}
In nature, neutrinos come in three flavors. However current experimental data indicate that the mass squared difference which dominates the oscillation of atmospheric neutrinos is much larger than that which dominates the oscillation of solar neutrinos~\cite{ParticleDataGroup:2022pth}. This allows us to reduce the problem of oscillation between the three flavors of neutrinos to an effective two-neutrino oscillation between $\nu_e$ and $\nu_x$, where $\nu_x$ is some linear combination of the two other flavors of neutrinos $\nu_{\mu}, \nu_{\tau}$~\cite{PhysRevD.74.123004}. We will track the flavor dynamics of neutrino oscillations in a  CCSN 
through the Liouville-von Neumann equation 
\begin{align}
    i\frac{\partial \rho}{\partial t}=[H,\rho]_{-}\,.
    \label{eq:LvN}
\end{align}
Here $\rho$ is a 2$\times$2 density matrix in flavor space, containing the information about the flavor content of the neutrino. The Hamiltonian can be split into three parts,
\begin{align}
    H=H_{V}+H_{M}+H_{\nu\nu}\,. \label{eq:components_of_H}
\end{align}
In this, $H_{V}$ is the part containing the contribution due to the mass mixing, which causes vacuum oscillations;
$H_M$ represents the contribution due to the presence of the non-neutrino matter, i.e. leptons and quarks; and
$H_{\nu\nu}$ corresponds to the self-interaction of neutrinos. In the two neutrino paradigm, we can work with a $2$-dimensional density matrix. We will denote the density matrix of neutrinos by $\rho$ and that of antineutrinos by $\bar{\rho}$. The general form of these matrices are
\begin{align}
\rho&=\frac{1}{2}n(\mathbb{I}_2+\vec{P}\cdot \vec{\sigma}), \label{eq:P_vector1} \\
\bar{\rho}&=\frac{1}{2}\bar{n}(\mathbb{I}_2+\vec{\bar{P}}\cdot \vec{\sigma})\,. \label{eq:Pbar_vector1}
\end{align}
The vector $\vec{P}$ and $\vec{\bar{P}}$ are called the polarization vector in literature. We will call it by the same name in this work. For example, consider a neutrino state of the form
\begin{align}
    \ket{\psi}=a_e\ket{\nu_e}+a_{x}\ket{\nu_{x}}
,\end{align}
which can be written as the 2$\times$2 density matrix
\begin{align}
    n\begin{pmatrix}
        |a_e|^2 & a_ea^{*}_{x} \\
        a^{*}_ea_{x} & |a_{x}|^2
    \end{pmatrix}=n\biggl(\frac{1}{2}\mathbb{I}+\frac{1}{2}\vec{P}\cdot\vec{\sigma}\biggr)\,,
\end{align}
so we can use $n$ to denote the total density of neutrinos. The components of the polarization vector $\vec{P}$ are then $(\Real(2a_e^{*}a_{x}),\Imag(2a_e^{*}a_{x}),|a_{e}|^2-|a_{x}|^2)$. The density matrix of the antineutrinos is defined analogously, with $\bar{n}$ the total number density of antineutrinos.

The vacuum mixing part of the Hamiltonian $H_V$ is (up to a term proportional to identity)
\begin{align}
H_V =& \frac{\Delta m^2}{4E}\begin{pmatrix} -\cos 2\theta & \sin 2\theta \\ \sin 2\theta & \cos 2\theta \end{pmatrix} \nn \\
=& \frac{\Delta m^2}{2E}\frac{1}{2}\vec{B}\cdot \vec{\sigma}\,, 
\label{eq:H_V} 
\end{align}
where we have defined $~\vec{B}=(\sin 2\theta,0,-\cos 2\theta)\,.$
{The mixing matrix $\mathcal{U}$ is}
\begin{align}
\mathcal{U}=\begin{pmatrix}\cos\theta & \sin\theta \\ 
-\sin\theta & \cos\theta \end{pmatrix}
.\end{align}
$H_M$ contains the contribution from weak interactions, i.e. the Wolfenstein potential~\cite{PhysRevD.17.2369} and the contribution from spin-torsion interaction with the background~\cite{Ghose:2023ttq}. The contribution from the spin-torsion interaction needs to be rotated by $\mathcal{U}$ to the flavor space. In our discussion, we will assume the spin-torsion interaction to be maximally parity violating purely left chiral ($\lambda^{i}_{R}=0$ in Eq.~\eqref{eq:spin-torsion_interaction}). Considering the flavor evolution due to the non-neutrino background, we can take the background average of the relevant part of the interaction Lagrangian in Eq. \eqref{eq:spin-torsion_interaction}. The background average Hamiltonian (matter potential) can be calculated from finite temperature field theory or from forward scattering~\cite{Ghose:2023ttq, Barick:2023wxx, Pal:1989xs, PhysRevD.17.2369},
\begin{align}
H_M =& \frac{\Delta \lambda \lambda_f n_f}{2\sqrt{2}}\begin{pmatrix} -\cos 2\theta & \sin 2\theta \\ \sin 2\theta & \cos 2\theta \end{pmatrix}\pm\sqrt{2}G_Fn_e\begin{pmatrix}1 & 0 \\ 0 & 0\end{pmatrix} \notag\\
=& \frac{\Delta \lambda \lambda_f n_f}{\sqrt{2}}\frac{1}{2}\vec{B}\cdot\vec{\sigma}\pm\sqrt{2}G_Fn_e\frac{1}{2}\vec{L}\cdot \vec{\sigma}\,. \label{eq:matter_lagrangian}
\end{align}
Here  $\lambda_{1,2}$ are the coupling constants that appear in the geometrical interaction Lagrangian from Eq.~\eqref{eq:spin-torsion_interaction} in the left-chiral part of the neutrino current,\, $\Delta \lambda = \lambda_2-\lambda_1$\,, and $\vec{L}=(0,0,1)$. The upper sign in  Eq.~\eqref{eq:matter_lagrangian} corresponds to neutrinos and the lower sign corresponds to antineutrinos. We have also defined $\lambda_f=\sum_d\lambda_d n_d/\sum_d n_d$\,, where $\lambda_d$ is the coupling constant with which the fermion current of species $d$ appears in the contortion (except neutrinos). Electron and total non-neutrino fermion density are represented by $n_e$ and $n_f$ respectively. The contributions only refer to the forward scattering of neutrinos.

The self-interaction Hamiltonian $H_{\nu\nu}$ is derived in Appendix~\ref{app:hamiltonian} for the case when all neutrinos are created with the same energy. $H_{\nu\nu}$ has two components ---  the part due to weak interactions is given by
(Eq.~\eqref{eq:self_weak_single_mode}),
\begin{align}
H_{\nu\nu}^W=\sqrt{2}G_F(\rho-\bar{\rho})\,,\label{eq:self_interaction_weak}
\end{align}
and there is also the geometrical interaction term $H_{\nu\nu}^{S}$ from Eq.~\eqref{eq:self_torsion_single_mode}. To write the explicit form of $H_{\nu\nu}^{S}$ it is convenient to use a matrix $\tilde{\Lambda}$\,, 
\begin{align}
\tL=\frac{\lT}{2}+\frac{\Delta\lambda}{2}\begin{pmatrix}-\cos 2\theta & \sin 2\theta \\ \sin 2\theta & \cos 2\theta\end{pmatrix}=\frac{\lT}{2}\mathbb{I}+\frac{\Delta \lambda}{2}\vec{B}\cdot\vec{\sigma}\,, \label{eq:torsion_matrix}
\end{align}
where we have defined $\lambda_{\operatorname{tot}}=\lambda_1+\lambda_2$. The first part of the self-interaction potential {in \eqref{eq:self_torsion_single_mode}} will have the form
\begin{align}
\operatorname{Tr}(\tL(\rho-\bar{\rho}))\tL=\frac{1}{2}\Delta \lambda^2 \vec{B}\cdot(n\vec{P}-\bar{n}\vec{\bar{P}})\vec{B}\cdot\vec{\sigma} \label{eq:first_term_nunu_torsion}
.\end{align}
The second term of Eq.~(\ref{eq:self_torsion_single_mode}) can be written as 
\begin{align}
\tL(\rho-\bar{\rho})\tL &=\frac{1}{2}(\lT+\Delta \lambda \vec{B}\cdot\vec{\sigma})\frac{1}{2}(n\vec{P}-\bar{n}\vec{\bar{P}})\cdot\vec{\sigma}\frac{1}{2}(\lT+\Delta \lambda \vec{B}\cdot\vec{\sigma}) \nn \\
&=\frac{1}{8}\left[\lT^2(n\vec{P}-\bar{n}\vec{\bar{P}})\cdot\vec{\sigma}
+ \lT\Delta\lambda(n\vec{P}-\bar{n}\vec{\bar{P}})\cdot\vec{B}
+\Delta \lambda^2\vec{B}\cdot(n\vec{P}-\bar{n}\vec{\bar{P}})\vec{B}\cdot\vec{\sigma}\right.\nn \\
&\qquad \left. -\Delta \lambda^2(n\vec{P}-\bar{n}\vec{\bar{P}})\cdot\vec{\sigma} 
-\Delta\lambda^2\vec{B}\cdot(n\vec{P}-\bar{n}\vec{\bar{P}})\vec{B}\cdot\vec{\sigma}\right] \nn \\
&=\frac{1}{8}(\lT^2-\Delta \lambda^2)(n\vec{P}-\bar{n}\vec{\bar{P}})\cdot\vec{\sigma}\,,
\end{align}
where we have used $|\vec{B}|^2 = 1$\,.
In the last step, we have neglected a term as it is proportional to identity and will not contribute to the commutator on the RHS of the Liouville-von Neumann equation. Hence, collecting all the terms we write the Hamiltonian from~\eqref{eq:self_torsion_single_mode} for a mono-energetic beam of neutrinos,
\begin{align}
H_{\nu\nu}^{S}=\frac{\sqrt{2}}{4}\frac{1}{2}[\Delta\lambda^2\vec{B}\cdot(n\vec{P}-\bar{n}\vec{\bar{P}})\vec{B}\cdot\vec{\sigma}+\frac{1}{4}(\lT^2-\Delta \lambda^2)(n\vec{P}-n\vec{\bar{P}})\cdot\vec{\sigma}]\,.\label{HnnS}
\end{align}
Remembering the commutation relation $[\sigma_i,\sigma_j]_{-}=2i\epsilon_{ijk}\sigma_k$\,, we calculate from Eq.~(\ref{eq:LvN}) the equations of motion for $\vec{P}$ and $\vec{\bar{P}}$\,,
\begin{align}
\partial_{t}\vec{P}=&\biggl[\frac{\Delta m^2}{2E}\hat{\omega}\vec{B}+\frac{\Delta \lambda \lambda_f n_f}{\sqrt{2}}\vec{B}+\frac{\sqrt{2}}{4}\Delta\lambda^2\vec{B}\cdot(n\vec{P}-\bar{n}\vec{\bar{P}})\vec{B}+\sqrt{2}G_Fn_e\vec{L}\nn \\
& \qquad +  \frac{\sqrt{2}}{4}\frac{1}{4}(\lT^2-\Delta \lambda^2|\vec{B}|^2)(n\vec{P}-\bar{n}\vec{\bar{P}}) +\sqrt{2}G_F(n\vec{P}-\bar{n}\vec{\bar{P}})\biggr]\times\vec{P}\,, \label{eq:P_eqn}
\end{align}

\begin{align}
\partial_{t}\vec{\bar{P}}=&\biggl[-\frac{\Delta m^2}{2E}\hat{\omega}\vec{B}+\frac{\Delta \lambda \lambda_f n_f}{\sqrt{2}}\vec{B}+\frac{\sqrt{2}}{4}\Delta\lambda^2\vec{B}\cdot(n\vec{P}-\bar{n}\vec{\bar{P}})\vec{B}+\sqrt{2}G_Fn_e\vec{L}
 \nn \\
& \qquad +\frac{\sqrt{2}}{4}\frac{1}{4}(\lT^2-\Delta \lambda^2|\vec{B}|^2)(n\vec{P}-\bar{n}\vec{\bar{P}})+\sqrt{2}G_F(n\vec{P}-\bar{n}\vec{\bar{P}})\biggr]\times\vec{\bar{P}}\,. \label{eq:Pbar_eqn}
\end{align}
Here we have defined $\hat{\omega}$, which parametrizes the mass hierarchy --- $\hat{\omega}$ takes the value $+1$ for normal  hierarchy (NH) and $-1$ for inverted hierarchy (IH). {In the framework of the chiral torsion that we introduced,} torsion is a completely non-dynamic field with no natural mass scale associated. Hence, the size of the coupling constants cannot be fixed from purely theoretical considerations. However, we can try to see the consequences of this interaction by choosing some reasonable values of these coupling constants. Accordingly, let us choose $\lambda_1^2=gG_{F}$\,, $\lambda_2=(2r+1)\lambda_1$\,, and $\lambda_f=a\lambda_1$\,, for some $g, r,$ and $a$ (we assume that $\lambda_1$ is non-vanishing). Assuming that the total density of neutrinos is the same as the total density of antineutrinos,  $n = \bar{n}\,,$ we divide Eq. \eqref{eq:P_eqn} and \eqref{eq:Pbar_eqn} by $\Delta m^2/(2E)$ to get the equations 
\begin{align}
\partial_{\tau}\vec{P}&=\biggl(\hat{\omega}\vec{B}+\sqrt{2}agr R_f\vec{B}+\sqrt{2}R_{\nu}gr^2\vec{B}\cdot(\vec{P}-\vec{\bar{P}})\vec{B}+\sqrt{2}R_e\vec{L}+\sqrt{2}R_{\nu}f_{g,r}(\vec{P}-\vec{\bar{P}})\biggr)\times\vec{P} \label{eq:P_eqn_red} \\
\partial_{\tau}\vec{\bar{P}}&=\biggl(-\hat{\omega}\vec{B}+\sqrt{2}agrR_f\vec{B}+\sqrt{2}R_{\nu}gr^2\vec{B}\cdot(\vec{P}-\vec{\bar{P}})\vec{B}+\sqrt{2}R_e\vec{L}+\sqrt{2}R_{\nu}f_{g,r}(\vec{P}-\vec{\bar{P}})\biggr)\times\vec{\bar{P}} \label{eq:Pbar_eqn_red}
.\end{align}
Here we have introduced the dimensionless time $\tau=\dfrac{t \Delta m^2}{2E}$\,, the dimensionless reduced density parameters $R_e=\dfrac{2G_Fn_eE}{\Delta m^2}$\,, $R_f=\dfrac{2G_Fn_fE}{\Delta m^2}$\,, and $R_{\nu}=\dfrac{2G_FnE}{\Delta m^2}$\,, and also written $f_{g,r}=1+\dfrac{g}{4}(2r+1)$. 
We need at least one $\lambda$ to be non-zero for it to affect the neutrino oscillation pattern. We choose it to be $\lambda_1$, then $\lambda_2$ can always be made to vanish by the choice $r=-0.5$.\\

It is easy to see that if we put $g=4$ and $r=0$  in Eq.~\eqref{HnnS}, the self-interaction $H_{\nu\nu}^{S}$ looks like
\begin{align}
    H^{S}_{\nu\nu}=\frac{\sqrt{2}}{4}\frac{1}{2}\frac{1}{4}n\lambda_{\operatorname{tot}}^2(\vec{P}-\vec{\bar{P}})\cdot \vec{\sigma}=\sqrt{2}G_{F}(\rho-\bar{\rho}) \label{eq:spin-to-weak}
.\end{align}
The Hamiltonian from spin-torsion interaction becomes identical to the weak universal coupling.

\section{Flavor evolution in uniform density}
\label{sec:uni_den}
After deriving the Eqs.~\eqref{eq:P_eqn_red} and~\eqref{eq:Pbar_eqn_red} it would be insightful to plot the evolution of $\vec{P}$ and $\vec{\bar{P}}$\,. As general closed-form solutions of the two coupled vector differential equations are not available, we will plot the numerical solutions of the Eqs.~\eqref{eq:P_eqn_red} and~\eqref{eq:Pbar_eqn_red}. In this section, we will work with uniform density.

\begin{figure}[hbtp]
     \centering
     \begin{subfigure}[b]{0.46\textwidth}
         \centering
         \includegraphics[width=\textwidth]{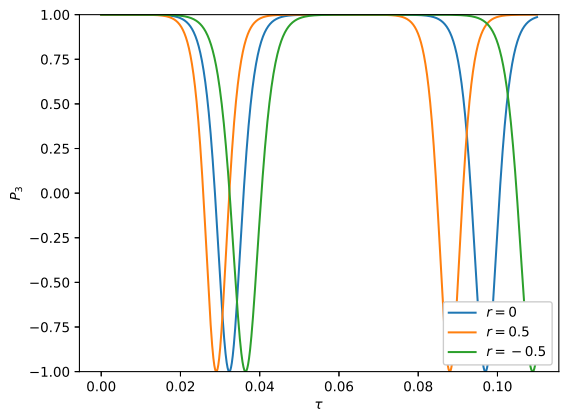}
         \caption{Evolution of $P_3$ with $a=0$ for different values of $r=0,0.5,-0.5$.}
         \label{fig:IH_first}
     \end{subfigure}
     \hfill
     \begin{subfigure}[b]{0.46\textwidth}
         \centering
         \includegraphics[width=\textwidth]{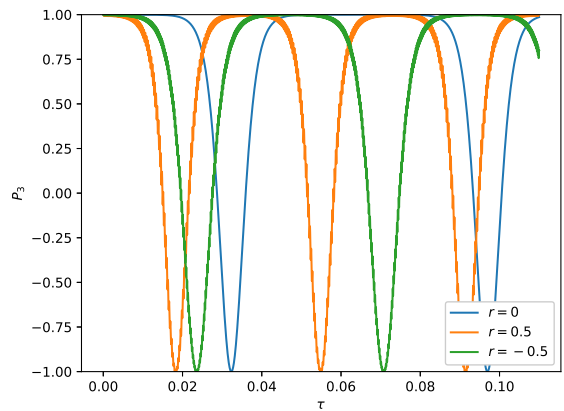}
         \caption{Evolution of $P_3$ with $a = 0.1$ for different values of $r=0,0.5,-0.5$.}
         \label{fig:IH_second}
     \end{subfigure}
        \caption{Shift in oscillation pattern in the presence of spacetime torsion with $g=1$.}
        \label{fig:firstgraphs}
\end{figure}

When the flavor dynamics of neutrinos are studied in the literature, the presence of ordinary matter is taken care of by making the effective vacuum mixing angle very small. The geometric interaction cannot be absorbed in the vacuum mixing angle in a similar manner. Hence, we will work with experimentally known values of the relevant vacuum mixing angle and a nonzero matter density. For purposes of illustration we use the neutrino oscillation parameters and relevant neutrino and matter densities as given in~\cite{Lin:2022dek}, then we are studying a system of monoenergetic neutrinos and antineutrinos with energy $E=15.1$ MeV. The mass squared difference is $\Delta m^2 = 2.5\times 10^{-3}$ eV$^2$\,, which  gives us a vacuum oscillation frequency $\Delta m^2/(2E)=12 \times 10^4$ s$^{-1}$\,, while the mixing angle is $8.6^\circ$. To express the densities of neutrinos and other relevant densities we introduce a number $\mu_0=1.76\times 10^5$, in units of which the reduced density parameters will be expressed. Let us now plot the neutrino flavor evolution in Fig.~\ref{fig:firstgraphs} for ($R_{\nu}$, $R_e$)=$(\mu_0/10,\mu_0/10)$. In our calculations, we have taken the densities of protons, neutrons, and electrons to be the same. Hence, for every electron, there will be 3 up quarks and 3 down quarks,
So we can write $R_f=7R_e$. In all these plots, the time axis will be marked as $\tau$\, which is time in units of $2E/\Delta m^2 = 8.3~\mu$s.

The bound on neutrino-neutrino self-interaction is still an area of active research~\cite{Cyr-Racine:2013jua, Camarena:2023cku, Das:2023npl}. All current bounds indicate that the neutrino-neutrino self-interaction coupling constant can be up to  6 or 7 orders of magnitude larger than in the Standard Model. We also remember
that $\lambda_1^2=gG_F$\, and that setting $g=4$ brings the strength of the geometric interaction to the same order as that of the Standard Model (see Eq.~\eqref{eq:spin-to-weak}). Hence, the term $f_{g,r}$ can deviate significantly from unity and we can get results that are very different from the standard model.

We plot the evolution of  $P_3$ with $\tau$  in Fig.~\ref{fig:firstgraphs} for the values of $(R_{\nu},R_{e})=(\mu_0/10,\mu_0/10)$. To understand the effects of different terms we at first only focus on the torsional contribution to the neutrino self-interaction part. For this purpose, we ignore the torsional contribution to the matter-neutrino interaction by setting $a=0$\,, but keep the usual Standard Model matter-neutrino interaction term.
 We have plotted the $P_3$ dynamics in this scenario in Fig.~\ref{fig:IH_first}. We have chosen the $r$ values such that the $\lambda_2$ is just as large as the SM. We see that positive values of $r$ increase the self-coupling and the oscillations start quicker, while negative values of $r$ decrease the self-interaction $R_{\nu}f_{g,r}$ and the oscillation starts slower. If we now choose a nonzero $a$, these terms contribute to the initial misalignment of the flavor states and this helps start the oscillation. We have plotted the dynamics of $P_3$ in such a scenario in Fig.~\ref{fig:IH_second}. It can be seen that the effect of non-zero $a$ is to bring the first minima closer to the origin compared to the $a=0$ patterns in the first panel. The matter potential induces small but rapid oscillations over the collective oscillation pattern, which adds to the thickness of the lines.

\begin{figure}[htbp]
     \centering
     \begin{subfigure}[b]{0.46\textwidth}
         \centering
         \includegraphics[width=\textwidth]{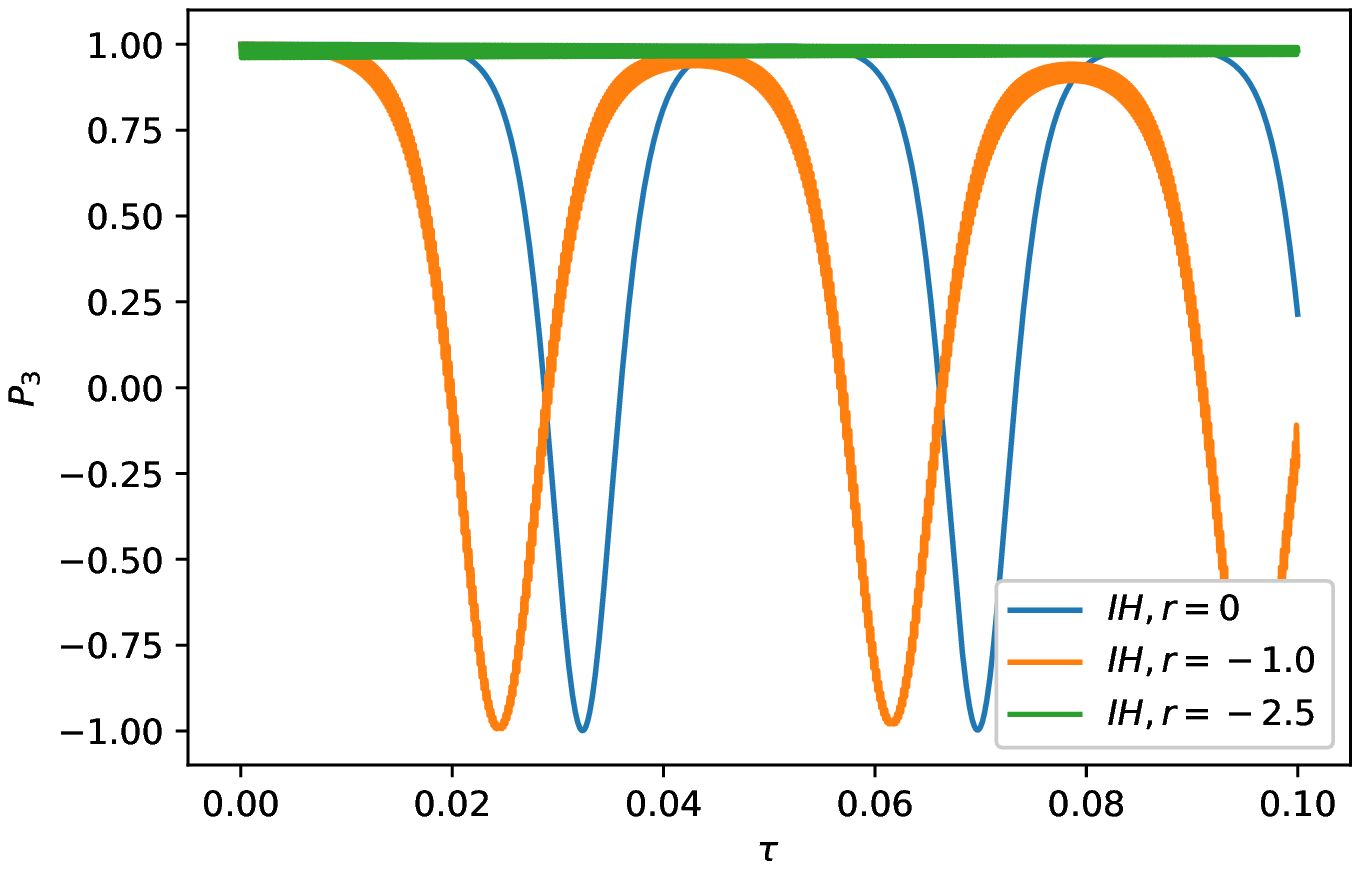}
         \caption{Evolution of $P_3$ in IH for larger values of $r$.}
         \label{fig:IH_suppressesd}
     \end{subfigure}
     \hfill
     \begin{subfigure}[b]{0.46\textwidth}
         \centering
         \includegraphics[width=\textwidth]{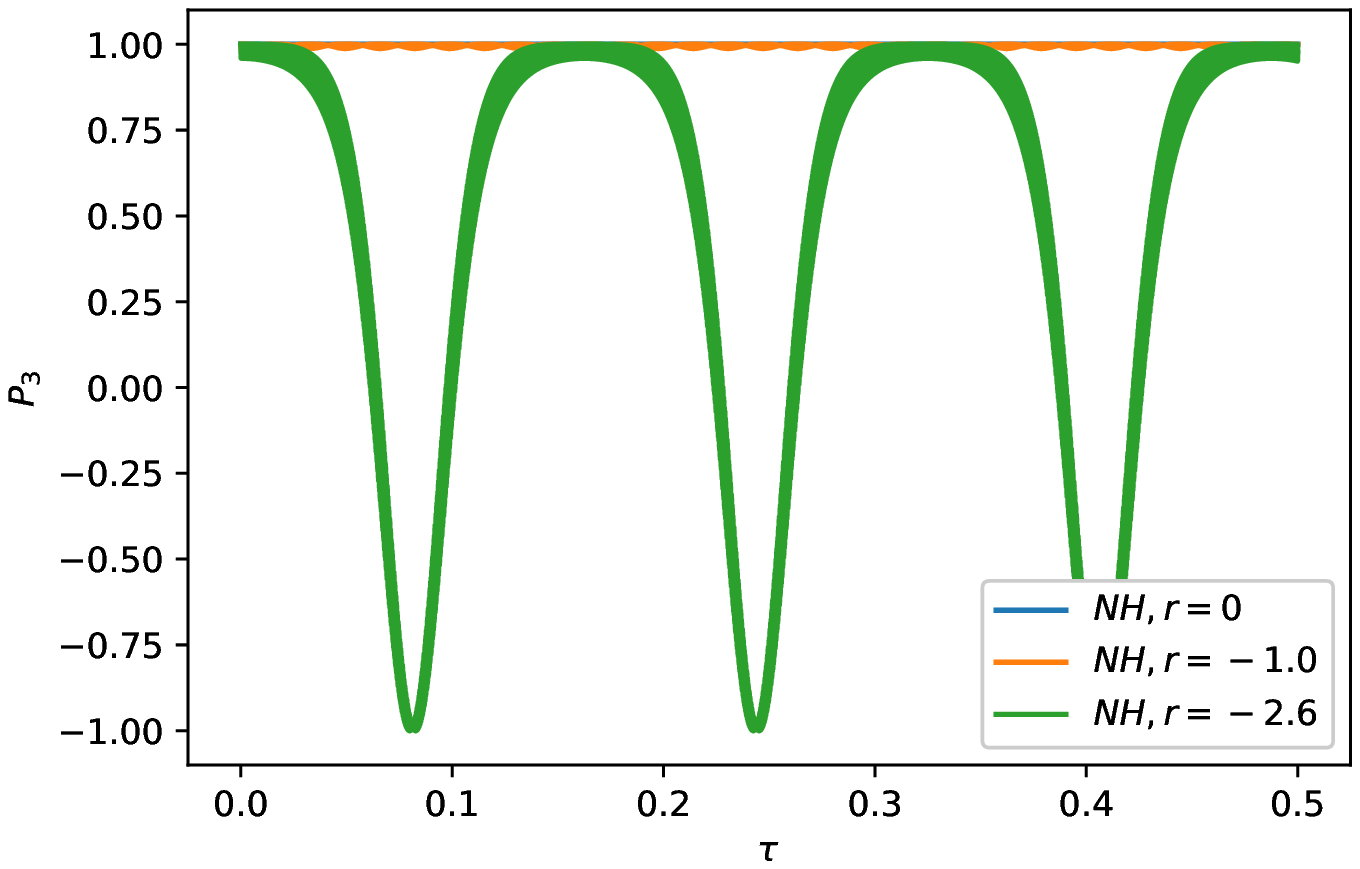}
         \caption{Evolution of $P_3$ in NH for larger values of $r$.}
         \label{fig:NH_suppressesd}
     \end{subfigure}
        \caption{Large negative values of $r$ can suppress flavor instability in IH and give rise to instability in NH. Both of these scenarios include $(R_{\nu},R_{e})=(\mu_0/10,\mu_0/10)$. We have chosen $g=1$ for the two plots.}
        \label{fig:secondgraphs}
\end{figure}

As already discussed, the bound on $r$ is quite loose. We can hence see the effects of such values of $r$ which makes the neutrino self-coupling constants larger than $G_F$ in both IH and NH. We have plotted two such cases in Fig.~\ref{fig:secondgraphs}. We have seen in the previous figures that positive values of $f_{g,r}$ IH give full flavor oscillation. However, the oscillation gets suppressed when $f_{g,r}$ starts becoming negative. This is what we have shown in Fig.~\ref{fig:IH_suppressesd}. As $r=-2.5$, $f_{g,r}=0$, and the oscillation gets suppressed quickly. In Fig.~\ref{fig:NH_suppressesd} we plot the same scenario but for NH. In this case positive values of $f_{g,r}$ don't cause any significant flavor conversion. However, a small but negative value of $f_{g,r}$ gives rise to flavor instability. The oscillation in NH builds up slowly. Hence, the time axis in Fig. \ref{fig:NH_suppressesd} is taken to longer. Thus if the coupling of the two neutrino mass eigenstates are of opposite sign and their difference is sufficiently large each hierarchy can show the flavor instability of the other hierarchy.

\begin{figure}[hbtp]
     \centering
     \begin{subfigure}[b]{0.46\textwidth}
         \centering
         \includegraphics[width=\textwidth]{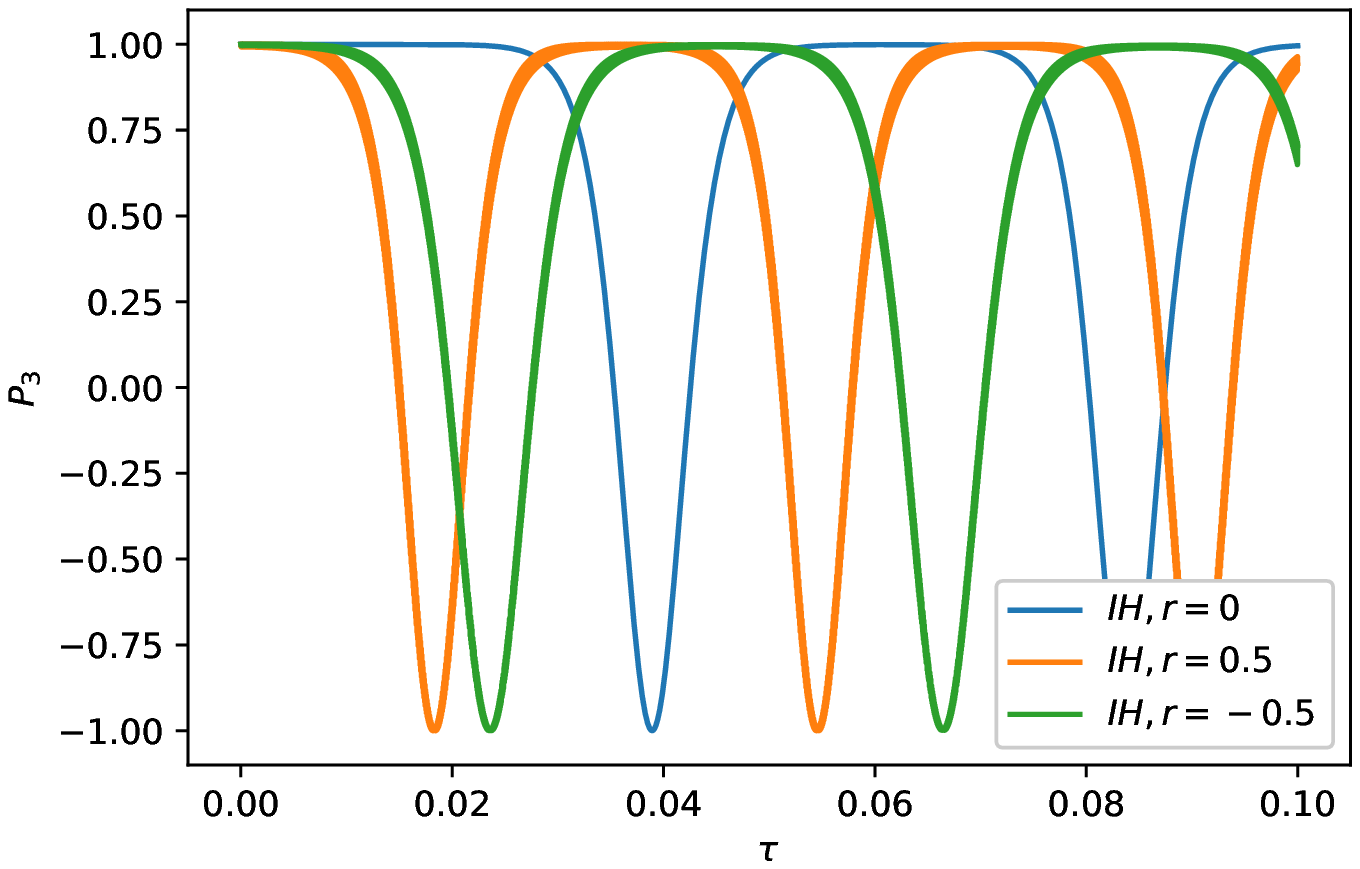}
         \caption{Evolution of $P_3$ for $(R_{\nu},R_e)=(\mu_0/10,\mu_0/2)$.}
         \label{fig:IHnext1}
     \end{subfigure}
     \hfill
     \begin{subfigure}[b]{0.46\textwidth}
         \centering
         \includegraphics[width=\textwidth]{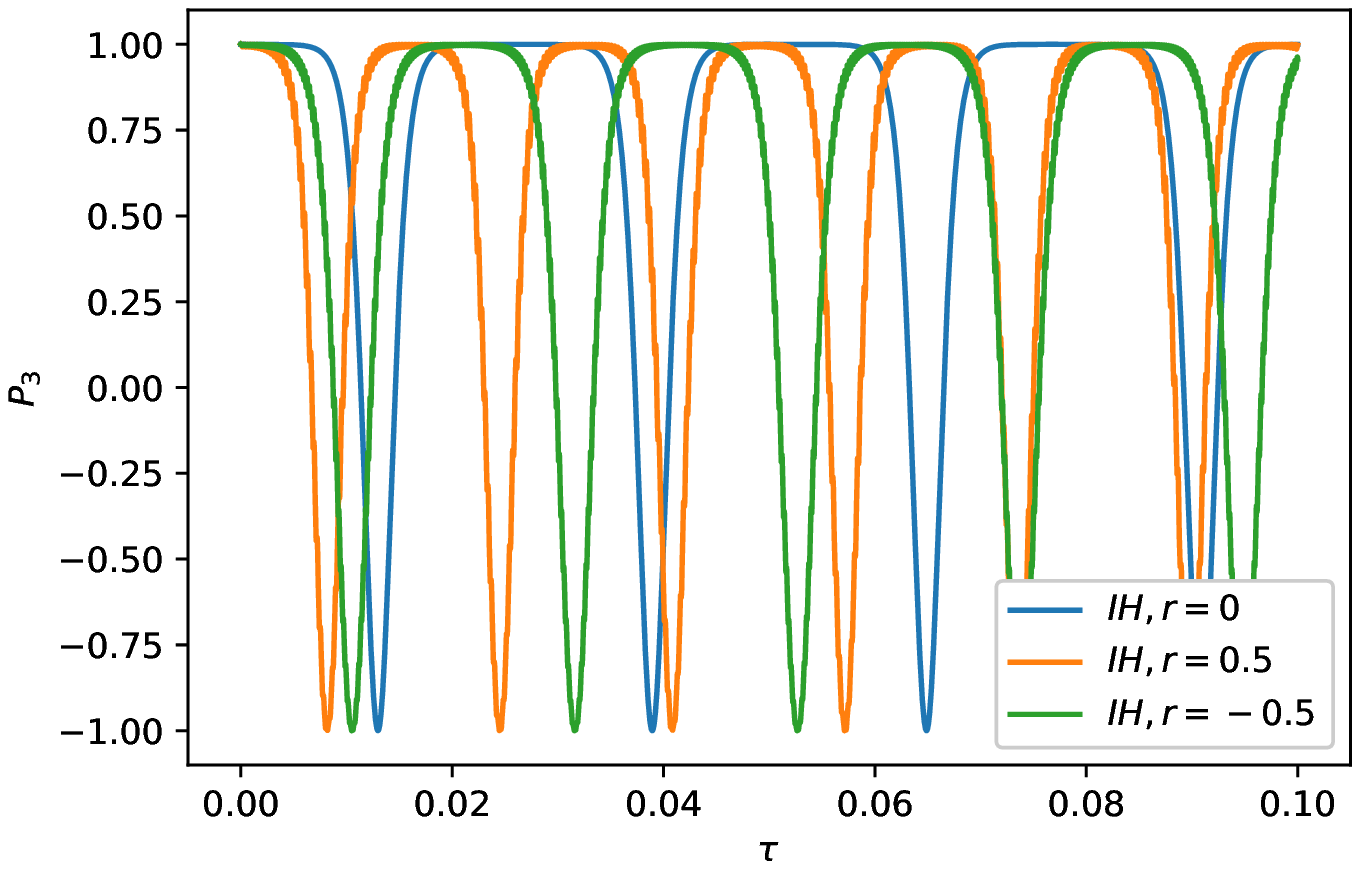}
         \caption{Evolution of $P_3$ for $(R_{\nu},R_e)=(\mu_0/2,\mu_0/10)$.}
         \label{fig:IHnext2}
     \end{subfigure}
        \caption{Evolution of $P_3$ when $R_{\nu}\neq R_{e}$. We have fixed $g=1$ in these.}
        \label{fig:thirdgraphs}
\end{figure}

The preceding discussion clears the contribution of the individual terms in the evolution of $P_3$ from Eqs.~\eqref{eq:P_eqn_red} and~\eqref{eq:Pbar_eqn_red}. Fig.~\ref{fig:thirdgraphs} plots two cases of $R_{\nu}\neq R_e$. In Fig.~\ref{fig:IHnext1} the higher values of $R_e$ make the plots thicker as the amplitude of small oscillations on top of the main oscillation patterns increases and the dip for $r=0$ gets shifted towards the right. This is consistent with the standard neutrino cosmology where the matter delays the onset of full flavor conversion. However, the effect of the spin-torsion coupling makes the dip move more to the left. This is a feature different from the standard neutrino oscillation picture. The same features are present in Fig.~\ref{fig:IHnext2}.  The higher value of $R_{\nu}$ increases the oscillation frequency and makes the oscillation pattern more rapid.

\section{Flavor evolution in a realistic supernova}
\label{sec:real_sn}

\begin{figure}[!ht]
     \centering
     \begin{subfigure}[t]{0.46\textwidth}
         \centering
         \includegraphics[width=\textwidth]{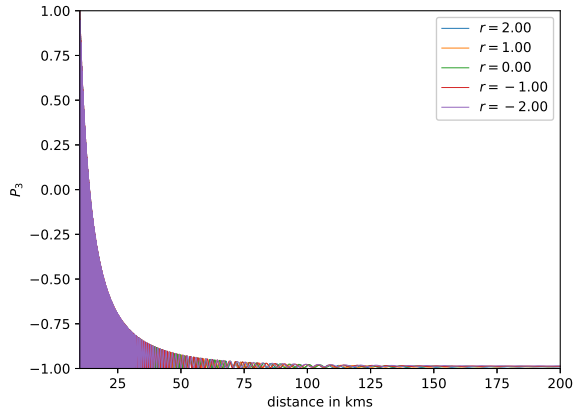}
         \caption{Evolution of $P_3$ for different values of $r$ when $g=0.5$}
         \label{fig:vary_a0_g_0.5}
     \end{subfigure}
     \hfill
     \begin{subfigure}[t]{0.46\textwidth}
         \centering
         \includegraphics[width=\textwidth]{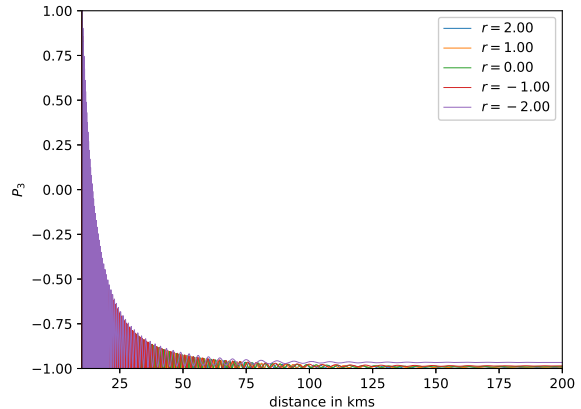}
         \caption{Evolution of $P_3$ for different values of $r$ when $g=1.5$}
         \label{fig:vary_a0_g_1.5}
     \end{subfigure}\\
     \begin{subfigure}[t]{0.46\textwidth}
         \centering
         \includegraphics[width=\textwidth]{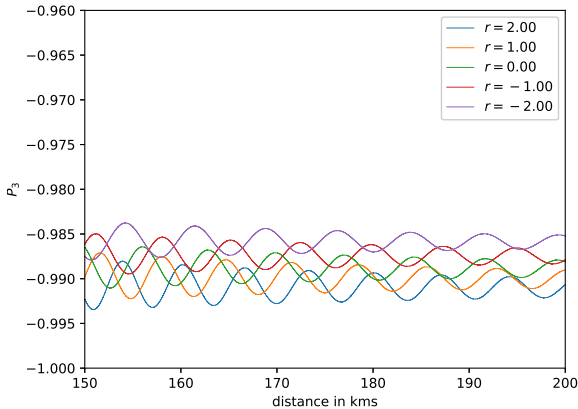}
         \caption{Detail of above figure}
         \label{fig:close-vary_a0_g_0.5}
     \end{subfigure}
     \hfill
     \begin{subfigure}[t]{0.46\textwidth}
         \centering
         \includegraphics[width=\textwidth]{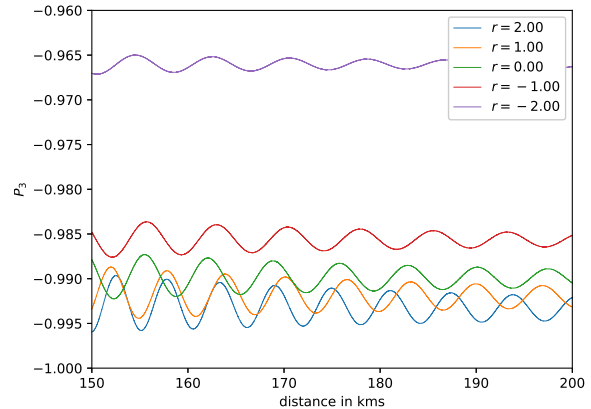}
         \caption{Detail of above figure}
         \label{fig:close-vary_a0_g_1.5}
     \end{subfigure}
        \caption{Evolution of $P_3$ for IH and 
        for uniform background density of fermions. In both panels, $a=0$ and $R_{\nu}(R)=R_{\bar{\nu}}(R)=\mu_0/10=R_e$. }
        \label{fig:vary_first}
\end{figure}

In Sec.~\ref{sec:uni_den}, we have only talked about flavor evolution in the presence of a uniform density of neutrinos. However, in a realistic supernova, the neutrino density falls as we go radially outward. In the presence of a uniform density of neutrinos, we saw that flavor evolution is periodic. That is to say, there is no permanent flavor evolution inside the supernova. The neutrinos move back and forth between the two flavors. Outside the core of a supernova, the neutrino flux falls off as the square of the propagation distance. If we also assume that the neutrinos are emitting semi-isotropically from the surface of the PNS and average over the angular distribution of neutrinos, the neutrino number density falls like~\cite{Duan:2006an}
\begin{align}
    R_{\nu,\bar{\nu}}(d)=R_{\nu,\bar{\nu}}(R)\left(1-\sqrt{1-\frac{R^2}{d^2}}\right)\frac{R^2}{d^2} \label{eq:neutrino_profile}
.\end{align}
$R$ is the radius of the proto-neutron star. We will be considering ultrarelativistic neutrinos, so the propagation distance is proportional to the time elapsed i.e. $d \propto t$. In terms of $\tau$, the neutrino density is
\begin{align}
    R_{\nu,\bar{\nu}}(\tau)=R_{\nu,\bar{\nu}}(R)\left(1-\sqrt{1-\frac{\tau_0^2}{\tau^2}}\right)\frac{\tau_0^2}{\tau^2} \label{eq:neutrino_density_time}
,\end{align}
where $\tau_0=R\Delta m^2/(2E)$. We will use the same values of $\theta,~\Delta m^2,~E$ as Sec.~\ref{sec:uni_den}. Then for $R=10$ km, we find $\tau_0=4$ for the neutrinos we are considering. Now, we will plot the oscillation patterns for different scenarios. 

\subsection{Uniform non-neutrino matter density $(a=0)$}
\label{sec:uniform_e_a_0}

\begin{figure}[!ht]
     \centering
     \begin{subfigure}[t]{0.46\textwidth}
         \centering
         \includegraphics[width=\textwidth]{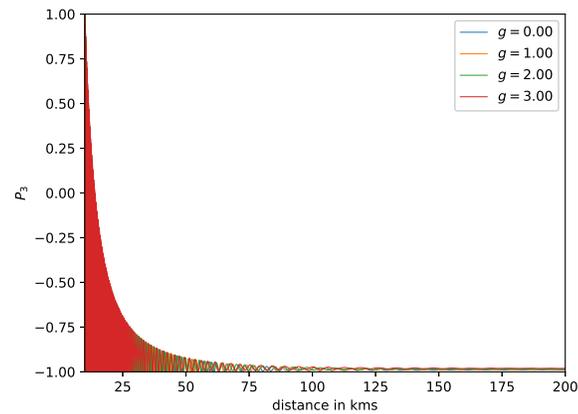}
         \caption{Evolution of $P_3$ when $r=-1.0$.}
         \label{fig:vary_a0_r_-1.0}
     \end{subfigure}
     \hfill
     \begin{subfigure}[t]{0.46\textwidth}
         \centering
         \includegraphics[width=\textwidth]{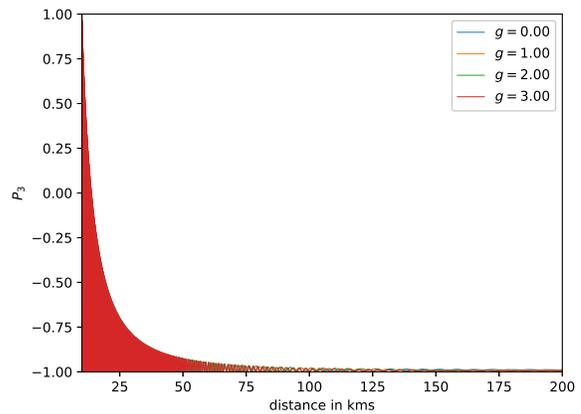}
         \caption{Evolution of $P_3$ when $r=+1.0$.}
         \label{fig:vary_a0_r_1.0}
     \end{subfigure}\\
     \begin{subfigure}[t]{0.46\textwidth}
         \centering
         \includegraphics[width=\textwidth]{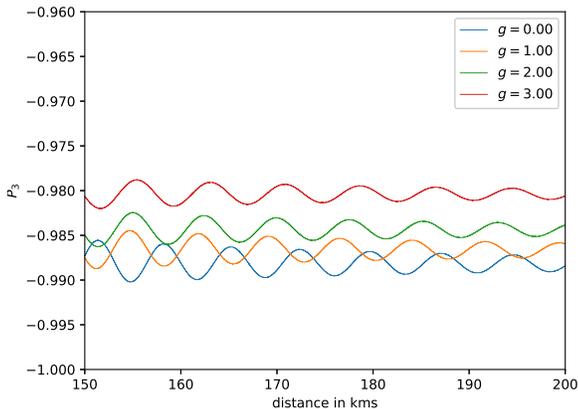}
         \caption{Detail of above figure}
         \label{fig:close-vary_a0_r_-1.0}
     \end{subfigure}
     \hfill
     \begin{subfigure}[t]{0.46\textwidth}
         \centering
         \includegraphics[width=\textwidth]{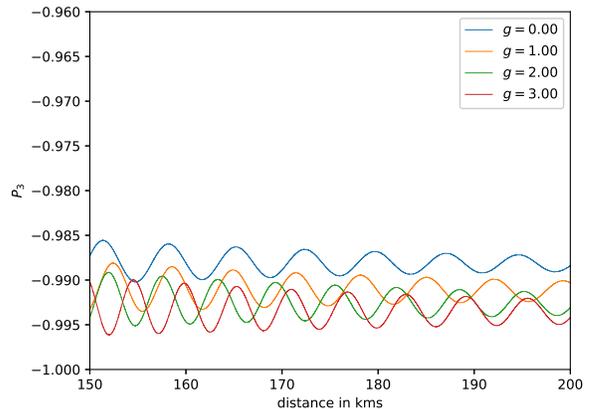}
         \caption{Detail of above figure}
         \label{fig:close-vary_a0_r_1.0}
     \end{subfigure}
        \caption{Evolution of $P_3$ for uniform background fermion density and for IH. In both panels, $a=0$ and $R_{\nu}(R)=R_{\bar{\nu}}(R)=\mu_0/10=R_e$.}
        \label{fig:vary_second}
\end{figure}

\begin{figure}[!ht]
     \centering
     \begin{subfigure}[t]{0.46\textwidth}
         \centering
         \includegraphics[width=\textwidth]{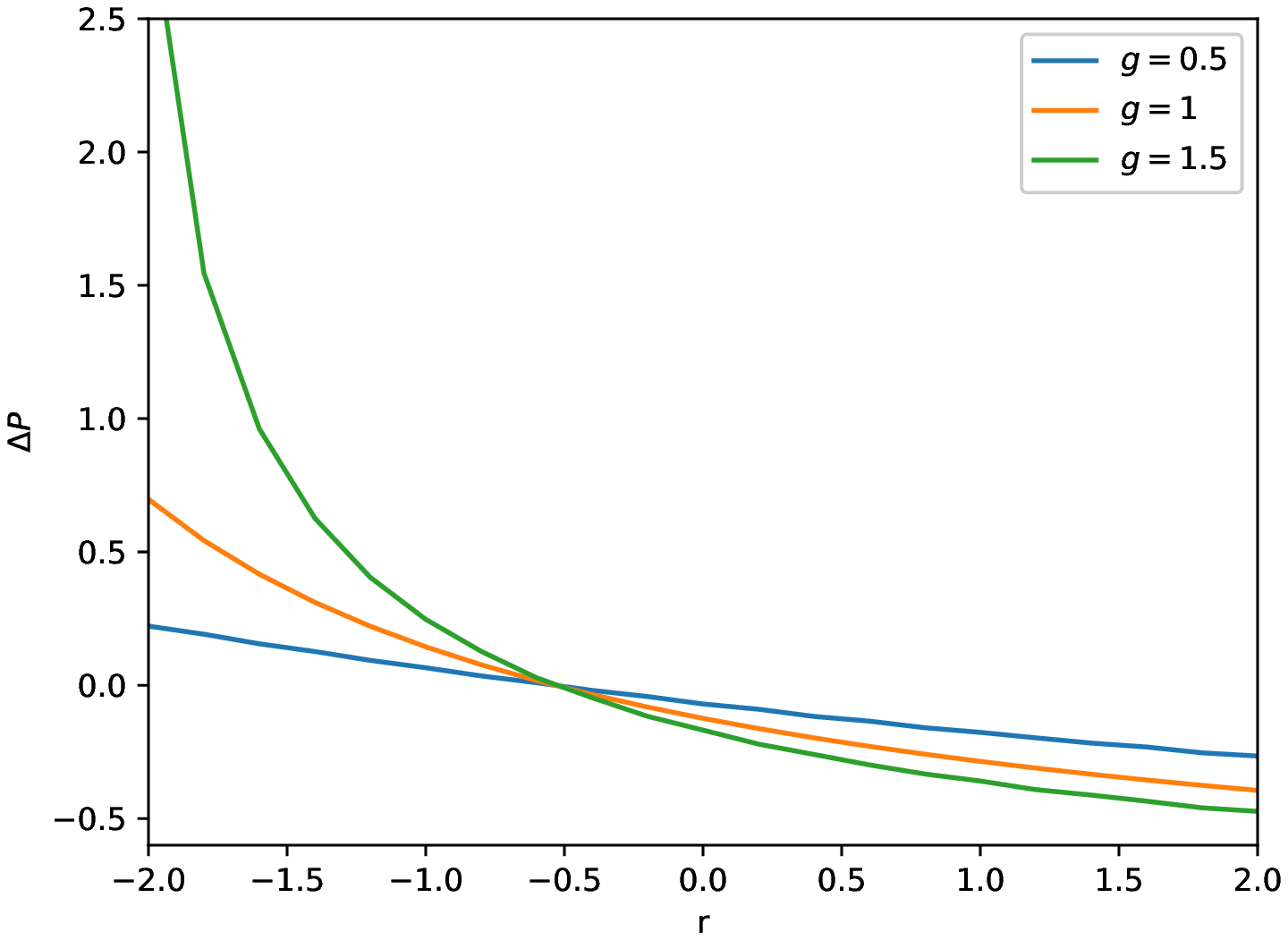}
         \caption{Fractional change in probability with varying $r$}
         \label{fig:g_fix_r_vary}
     \end{subfigure}
     \hfill
     \begin{subfigure}[t]{0.46\textwidth}
         \centering
         \includegraphics[width=\textwidth]{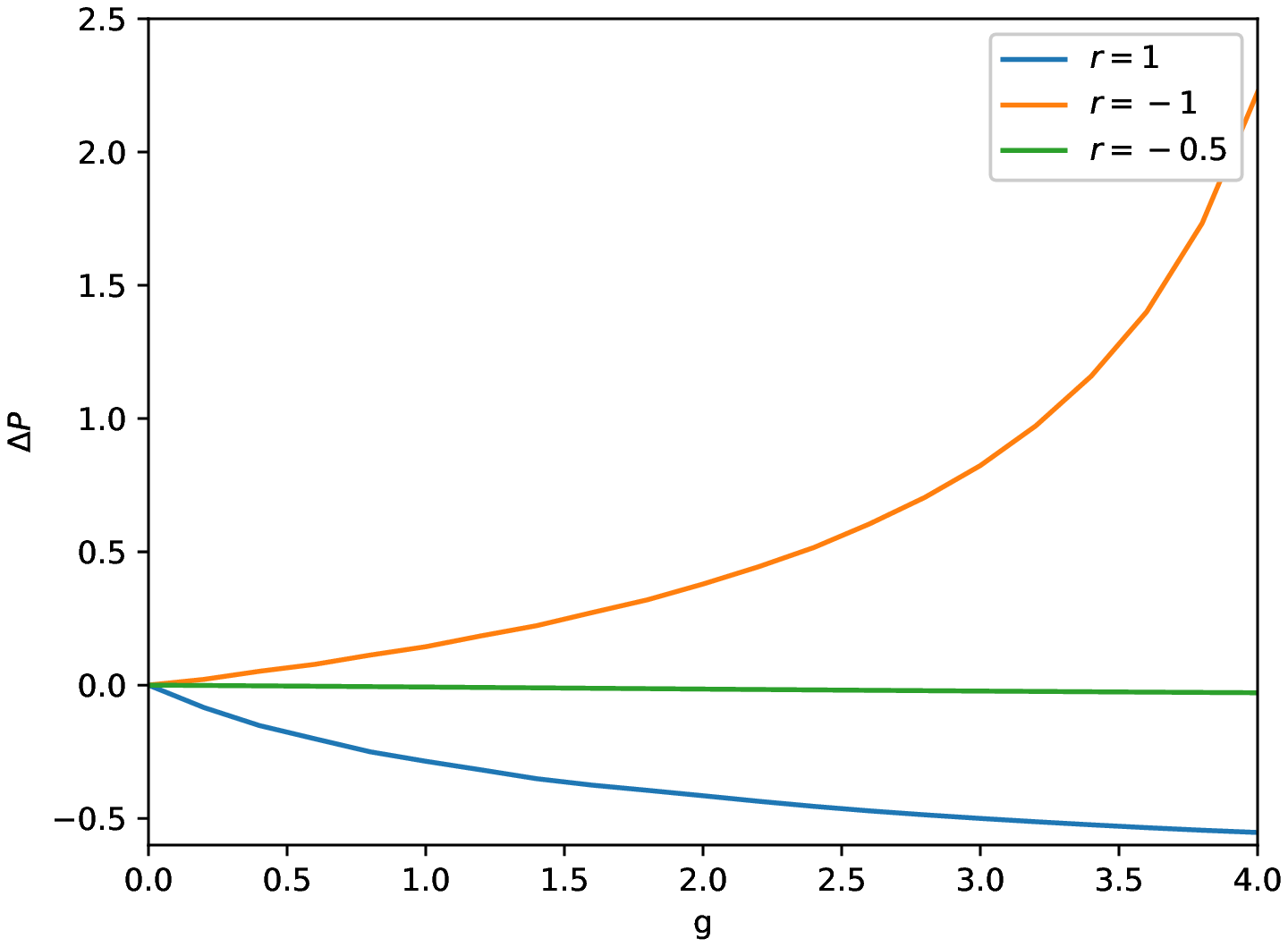}
         \caption{Fractional change in probability with varying $g$}
         \label{fig:g_vary_r_fix}
     \end{subfigure}
        \caption{Plot of $\Delta P$ for uniorm background fermion matter density and $a=0$.}
        \label{fig:frac_g_vary_r_vary}
\end{figure}
In this section non-neutrino matter density is uniform and the spin-torsion interaction with the non-neutrino matter is zero. The stability of solutions Eqs.~\eqref{eq:P_eqn_red} and~\eqref{eq:Pbar_eqn_red} requires very small step-size in the presence of variable fermion density. To choose the proper step size we compared solutions with a certain step size with the one calculated with a step size one order of magnitude smaller. When the solutions do not change by taking a smaller step size, we deduced that to be the proper step size for a good solution of the Eqs.~\eqref{eq:P_eqn_red} and~\eqref{eq:Pbar_eqn_red}. We saw that the solution by considering the step-size of $10^{-6}$ m does not change from $10^{-7}$ m. So, we considered $10^{-6}$ m to be a good choice for a stable solution.
{We have checked that if we assume normal mass hierarchy (NH), the density of neutrinos, and hence the geometrical self-interaction of neutrinos, 
fall off too quickly in this scenario to induce flavor instability. 
We will therefore focus only on the case of inverted mass hierarchy (IH) in the rest of the paper.}

{We plot in Figs.~\ref{fig:vary_first} and~\ref{fig:vary_second} the oscillation patterns for such a scenario when $(R_{\nu}(R),R_{e}(R))=(\mu_0/10,\mu_0/10)$\,.} In the presence of a uniform density of neutrinos, large enough negative values of $r$ can induce flavor oscillation to a neutrino gas of normal mass hierarchy as already shown in the previous section.
%
In Fig.~\ref{fig:vary_first}, several different values of $r$ are considered for fixed $g$\,, while in Fig.~\ref{fig:vary_second}  $g$ is varied for fixed $r$\,.
A generic feature of all the oscillation patterns is that as the neutrinos propagate away from the PNS, the electron survival probability falls off and settles to a stable value after propagation through around $100$ km. In all of the plots we have shown, all the neutrinos start as electron neutrinos and in all cases, most of the electron neutrinos convert to $\nu_{x}$. We see that the larger positive value of $r$ decreases the survival probability of the electron neutrinos and the larger negative value of $r$ increases the survival probability of electron neutrinos as long as $f_{g,r}$ is positive. As $f_{g,r}$ becomes negative, the oscillation dies down. 
{This feature is seen in~\ref{fig:close-vary_a0_g_0.5} and~\ref{fig:close-vary_a0_g_1.5}. This can be physically understood as the increasing $r$ increases the strength of the self-interaction and more electron neutrinos get converted into $\nu_{x}$. Also, the higher value of $g$ increases the strength of this self-interaction if $(2r+1)>0$, and the survival probability of the electron neutrino lowers slightly. The opposite happens for $2r+1 < 0$. This is seen in Figs.~\ref{fig:close-vary_a0_r_-1.0} and~\ref{fig:close-vary_a0_r_1.0}.} Increasing the value of $g$ increases the self-coupling and the survival probability decreases. In all of the cases, the electron neutrino survival probability falls and gets steady after propagating through $\approx 100$ km. Propagating the neutrino through $200$ km for different values of $g,r$ indicates clearly that the neutrino flavor oscillation has stopped. This is shown in Figs.~\ref{fig:vary_first} and~\ref{fig:vary_second}.

The initial oscillation of $P_3$ will not be visible in a terrestrial detector. The only value of $P_3$ relevant is the one that $P_3$ settles to after around $100$ km of propagation. Hence, we will study the survival probability changes with torsion parameters $g,r$ far away from the PNS. We will denote the value $P_3$ saturates as by $P_{\infty}(g,r)$, where the $g, r$ values denote at what values of the torsion parameters the $P_{\infty}$ was calculated. The survival probability of electron neutrinos far away from PNS ($\mathcal{P}_{S}$) is

\begin{align}
    \mathcal{P}_{S}=\frac{1}{n}\operatorname{Tr}\left(\rho \begin{pmatrix}
        1 & 0\\
        0 & 0
    \end{pmatrix}\right)=\frac{1+P_3}{2} \label{eq:survival_probability}
.\end{align}
Thus we find that the fractional change $\Delta P(g,r)$ in the probability will be
\begin{align}
    \Delta P(g,r)=\frac{\mathcal{P}_{S}(g,r)-\mathcal{P}_{S}(0,0)}{P_{S}(0,0)}=\frac{P_{\infty}(g,r)-P_{\infty}(0,0)}{1+P_{\infty}(0,0)} \label{def:del_P}
.\end{align}
We can now study the effects of the torsion parameter on $\Delta P$ by plotting them by varying the torsion parameters.

Figs.~\ref{fig:frac_g_vary_r_vary} summarizes the behavior of the electron neutrino survival probability with the torsion parameters. Increasing values of $r$ decrease the survival probability and negative values of $r$ quickly increase the survival probability for a given $g$. That is seen in Fig.~\ref{fig:g_fix_r_vary}. We see that for a positive value of $2r+1$ survival probability decreases with increasing $g$ and the opposite for a negative value of $2r+1$. Another interesting feature to note is that the survival probability away from PNS is independent of $g$ when $2r+1=0$. This shows that the effect of $f_{g,r}$ dominates all other $r$ dependent terms far away from PNS. For certain values of $(g,r)$ the fractional change in survival probability can be more than $2$ if the neutrino mass spectrum is inverted. As the number of events detected is proportional to the survival probability, the next-generation detectors can be used to probe the flavor contents of the supernova neutrino flux.

\subsection{Uniform non-neutrino matter density $(a \ne 0)$}
\label{sec:uniform_e_a_0.1}

\begin{figure}[!ht]
    \centering
    \begin{subfigure}[t]{0.46\linewidth}
        \includegraphics[width=\linewidth]{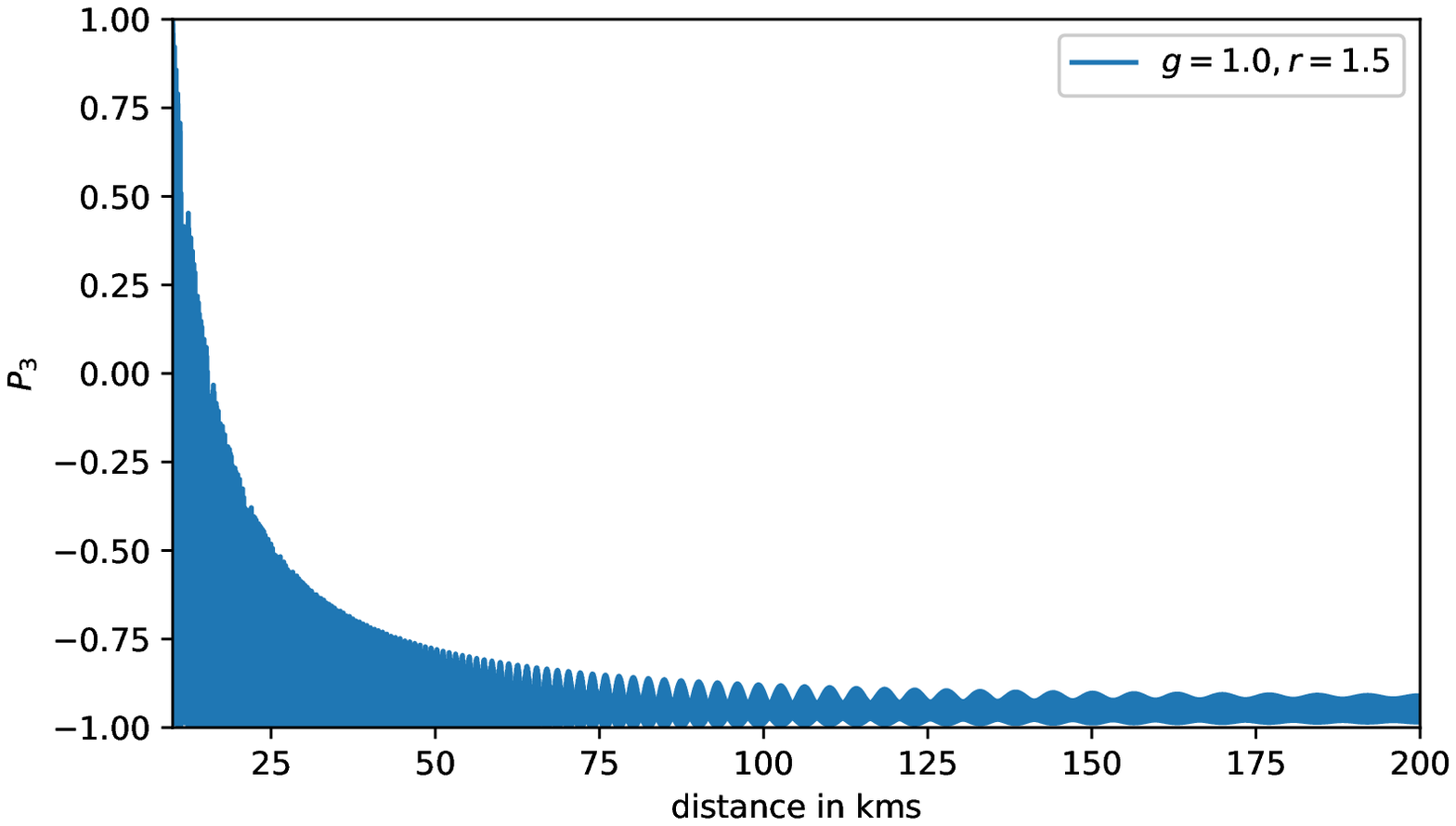}
        \caption{Oscillation pattern for $g = 1.0, r = 1.50$.}
        \label{fig:const-g_1.0_r_1.5_a_0.1}  
    \end{subfigure}
    \begin{subfigure}[t]{0.46\linewidth}
        \includegraphics[width=\linewidth]{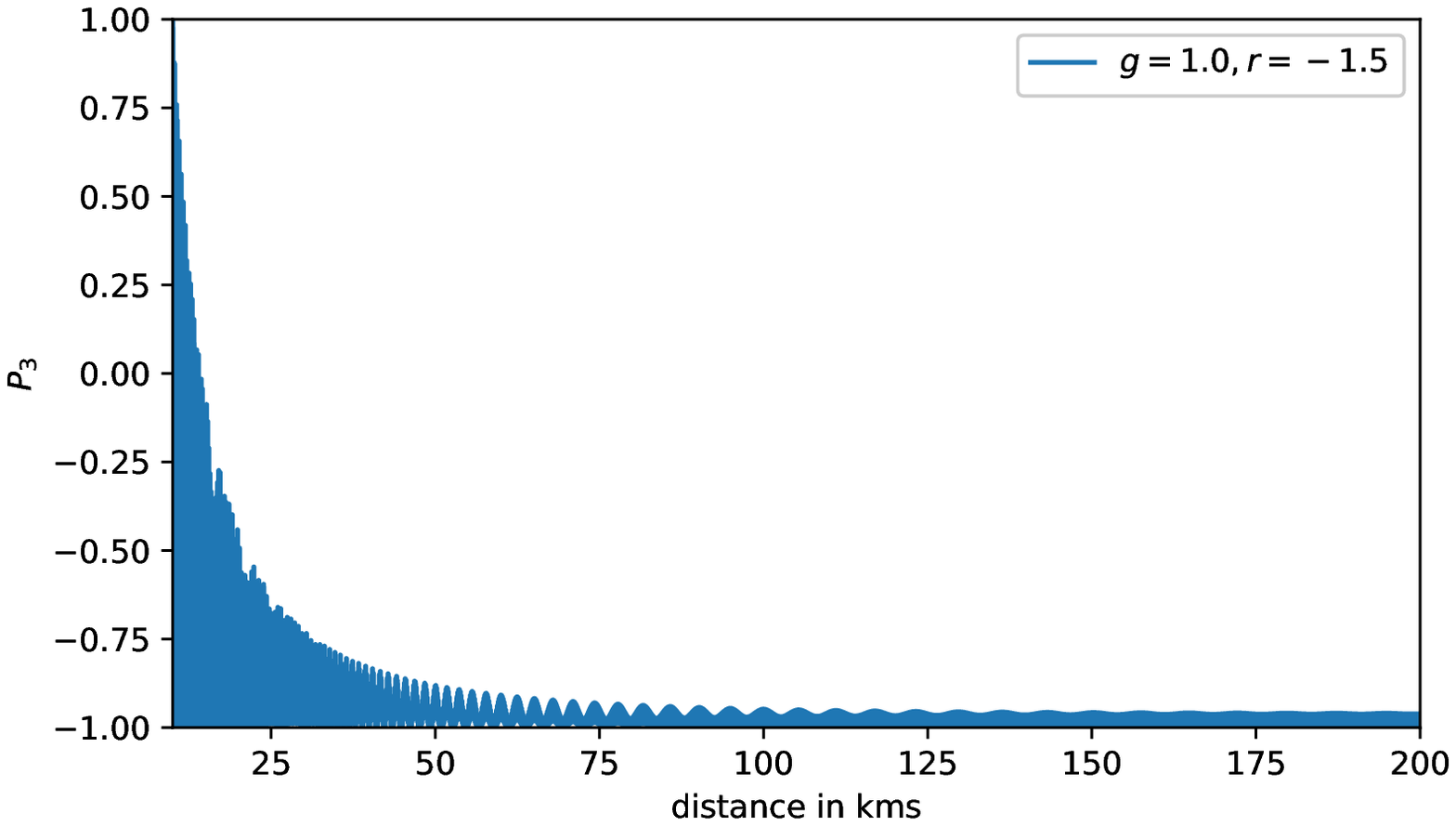}
        \caption{Oscillation pattern for $g = 1.0, r = -1.50$.}
        \label{fig:const-g_1.0_r_-1.5_a_0.1}  
    \end{subfigure}\\
    \begin{subfigure}[t]{0.46\linewidth}
        \includegraphics[width=\linewidth]{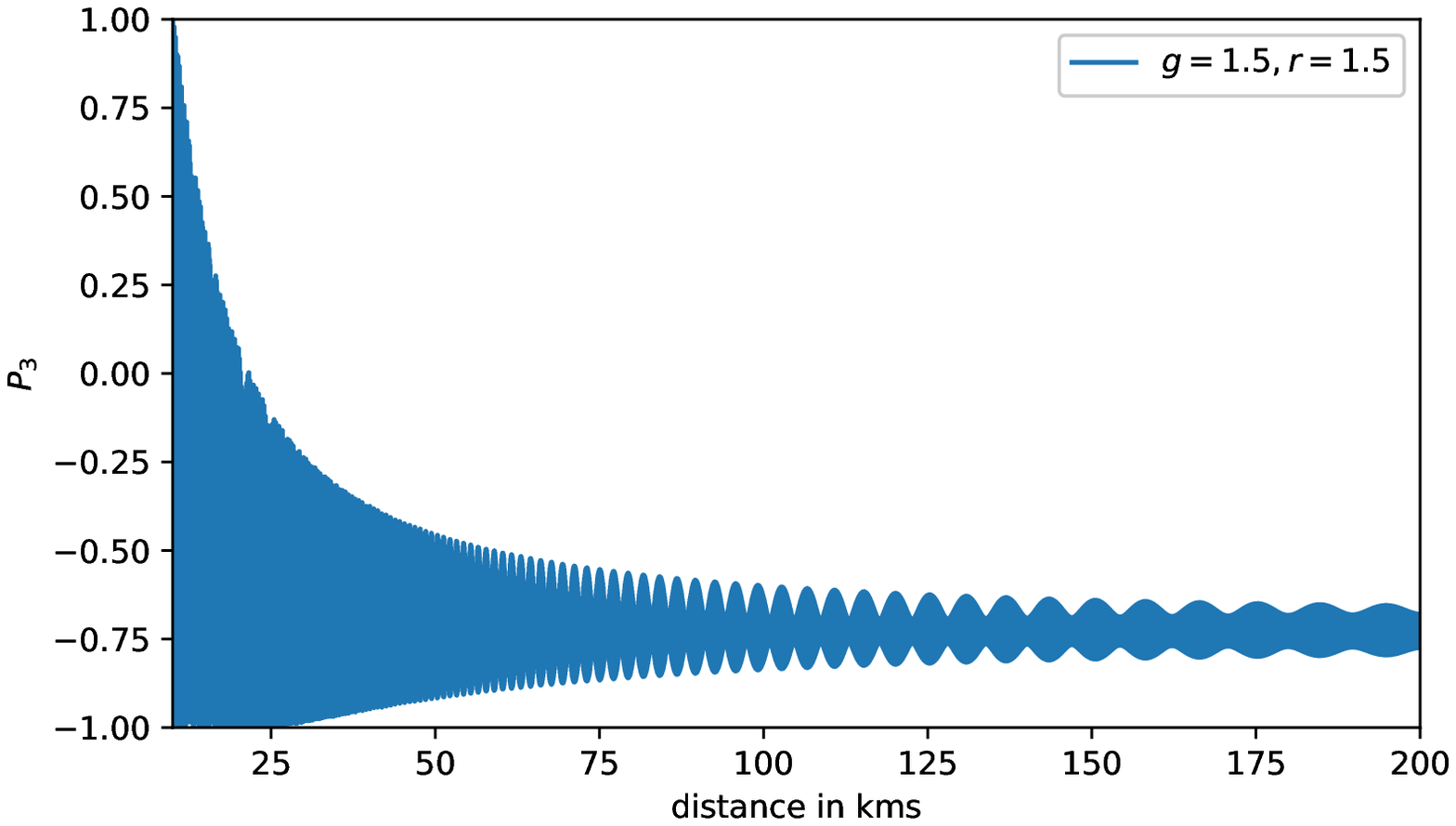}
        \caption{Oscillation pattern for $g = 1.5, r = 1.50$.}
        \label{fig:const-g_1.5_r_1.5_a_0.1}  
    \end{subfigure}
    \begin{subfigure}[t]{0.46\linewidth}
        \includegraphics[width=\linewidth]{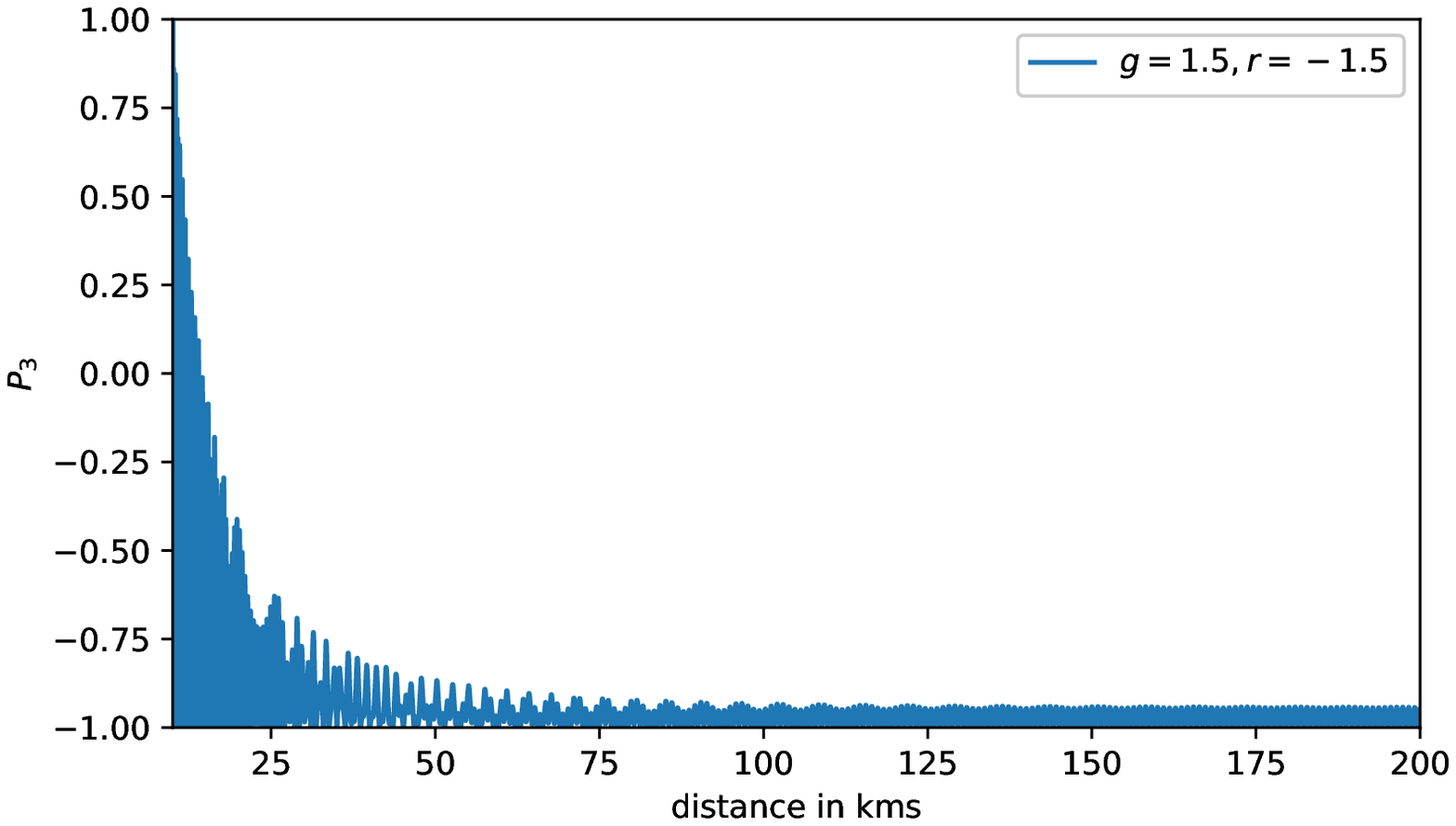}
        \caption{Oscillation pattern for $g = 1.5, r = -1.50$.}
        \label{fig:const-g_1.5_r_-1.5_a_0.1}  
    \end{subfigure}
    \caption{Oscillation patterns for uniform background fermion density and $a=0.1$.}
    \label{fig:const-a_0.1}
\end{figure}
\begin{figure}[!htp]
    \centering
    \begin{subfigure}[t]{0.46\linewidth}
        \includegraphics[width=\linewidth]{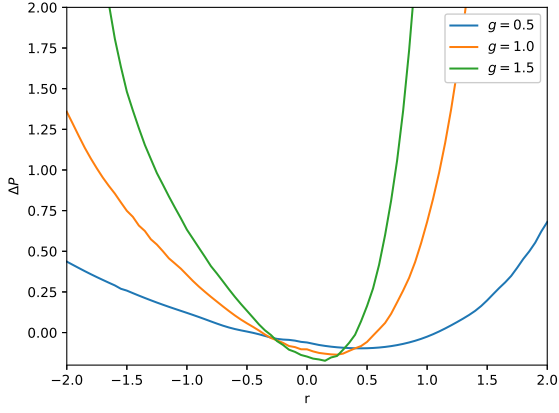}
        \caption{Fractional change in probability when varying $r$.}
        \label{fig:const-g_fix_a_0.1}
    \end{subfigure}
    \hfill
    \begin{subfigure}[t]{0.46\linewidth}
        \includegraphics[width=\linewidth]{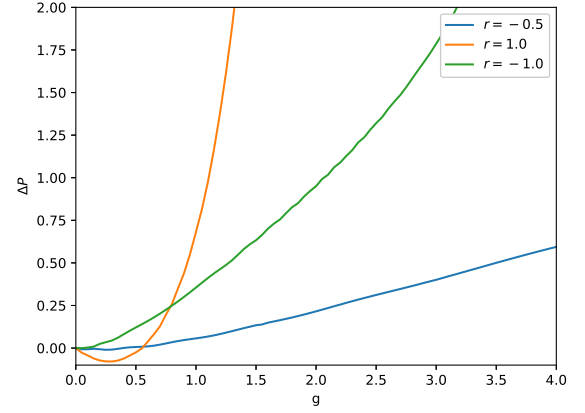}
        \caption{Fractional change in probability when varying $g$.}
        \label{fig:const-r_fix_a_0.1}
    \end{subfigure}
    \caption{Plot of $\Delta P$ for uniform background fermion density and $a=0.1$.}
    \label{fig:Delta_P_const_a_0.1}
\end{figure}
In the previous section, {we ignored the geometrical coupling between neutrinos and background matter by setting $a=0$\,. Then }
the presence of uniform non-neutrino matter density affects the flavor content of the neutrino flux coming out of a CCSN. In this section, we will consider a non-zero geometrical coupling of neutrinos with a background of uniform density. Throughout this section, we will set $a=0.1$, { which is neither too small nor too large}. We see the oscillation patterns in Fig.~\ref{fig:const-a_0.1}.

We still see a decrease in survival probability when the neutrinos go radially outwards. However, the interaction with the non-neutrino background keeps the flavors oscillating around a mean value instead of reaching an equilibrium we have seen in the previous section. Hence, in this case, the average value of $P_3$ over a large number of periods after about $100$ km is the quantity that will be used to calculate the fractional change in probability. Hence, we redefine $\Delta P$ as
\begin{align}
    \Delta P(g,r)=\frac{\braket{\mathcal{P}_{S}(g,r)}-\braket{\mathcal{P}_{S}(0,0)}}{\braket{P_{S}(0,0)}}=\frac{P_{\infty}(g,r)-P_{\infty}(0,0)}{1+P_{\infty}(0,0)} \label{def:del_P_average}
.\end{align}
Here the angular brackets represent averaging over a sufficiently large number of periods. The value $P_3$ settles into far away from the supernova's core is written as $P_{\infty}$ and is calculated by an average of $P_3$ over many periods. This means $P_{\infty} = \braket{P_3}$. To study the effects of the torsion parameters on the flux reaching a detector we will calculate $\Delta P$ again. $\Delta P$ is plotted by varying $g, r$ in Fig.~\ref{fig:Delta_P_const_a_0.1}.
It is seen from Fig.~\ref{fig:const-r_fix_a_0.1} that the fractional change in probability away from the core shows an approximate $r \to -r$ symmetry. The survival probability increases in both directions as we move away from the $r=0$ point. The rise in $\Delta P$ is quicker with higher $g$ as it increases the spin-torsion coupling. $\Delta P$ is also seen to be rising much quicker with higher values of $r$. With increasing strength of the spin-torsion interaction, the $\Delta P$ rises much quicker. 

\begin{figure}[!hbp]
    \centering
    \begin{subfigure}[b]{0.46\linewidth}
        \includegraphics[width=\linewidth]{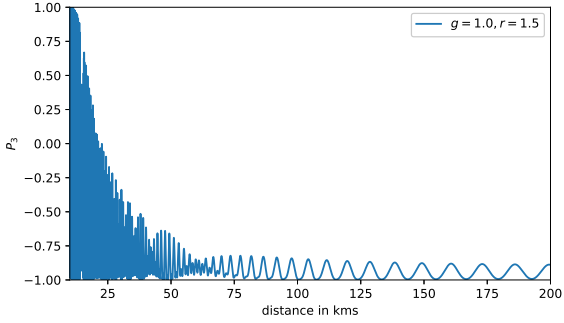}
        \caption{Oscillation pattern for $g = 1.0, r = 1.50$.}
        \label{fig:exp-g_1.0_r_1.5_a_0.1}  
    \end{subfigure}
    \hfill
    \begin{subfigure}[b]{0.46\linewidth}
        \includegraphics[width=\linewidth]{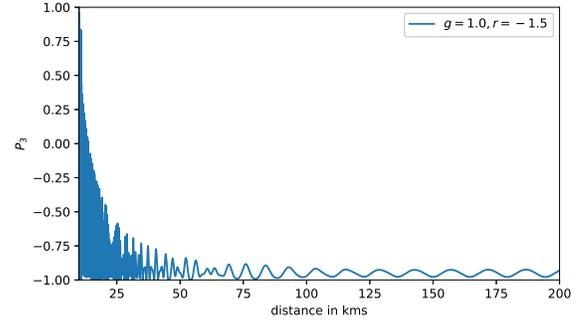}
        \caption{Oscillation pattern for $g = 1.0, r = -1.50$.}
        \label{fig:exp-g_1.0_r_-1.5_a_0.1}  
    \end{subfigure}\\
    \begin{subfigure}[b]{0.46\linewidth}
        \includegraphics[width=\linewidth]{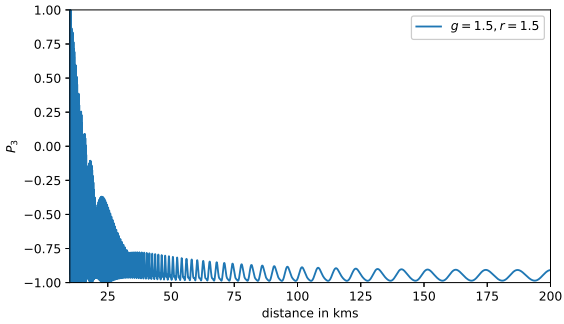}
        \caption{Oscillation pattern for $g = 1.5, r = 1.50$.}
        \label{fig:exp-g_1.5_r_1.5_a_0.1.eps}  
    \end{subfigure}
    \hfill
    \begin{subfigure}[b]{0.46\linewidth}
        \includegraphics[width=\linewidth]{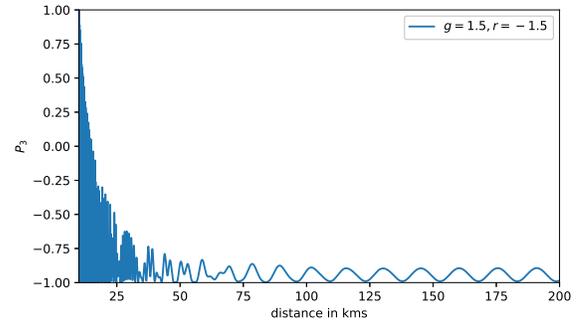}
        \caption{Oscillation pattern for $g = 1.5, r = -1.50$.}
        \label{fig:exp-g_1.5_r_-1.50_a_0.10}  
    \end{subfigure}
    \caption{Oscillation patterns when the background density decays exponentially and $a=0.1$.}
    \label{fig:exp-a_0.1}
\end{figure}

\subsection{Non-uniform density of non-neutrino matter (non zero a)}
\label{sec:non-uniform_e}

In Sec.~\ref{sec:uniform_e_a_0} and~\ref{sec:uniform_e_a_0.1} we have seen the asymptotic survival probability in the presence of uniform density of non-neutrino matter. In this section, we will consider a more realistic approach where the electron density falls off with distance. We will model the electron density as 
\begin{equation}
R_{e}(d)=\exp\left(-\frac{d-R}{0.18}\right) \label{eq:e_density}
\end{equation}
following~\cite{Duan:2006an}, where the $0.18$ factor is calculated assuming the $d$ is expressed in km. Here, $d$ is the propagation distance from the center of the PNS. As before $R_f=7R_e$ at every point. We will plot the oscillation patterns for two sample values of $g, r$ in Fig.~\ref{fig:exp-a_0.1}.  

Both the panels show a permanent decrease in survival probability as the neutrinos propagate radially outwards. In all the panels we see that away from the core the new oscillations are slower than the corresponding cases in uniform background fermion density in Figs.~\ref{fig:const-a_0.1}. It shows that the oscillation periods get longer in the presence of exponentially falling fermion density.

To study the effect of torsional couplings we will again calculate $\Delta P$ for different values of the parameters $g, r$. On calculating the $\Delta P$ for different values of $g, r$ we cannot see any obvious signatures of $g, r$ as opposed to what was seen in the case of uniform background fermion density for the values of $g, r$ considered. For the background fermions density profile considered, the density falls off by a factor $e^{-10}$ after about $2$ km. The oscillations that we see far away from the core are effectively vacuum oscillations. If the mass spectrum of the neutrinos would have been degenerate we would still see an oscillation due to four fermion torsional interaction but the new we would not see any vacuum oscillation in this case. The oscillation pattern would resemble Figs.~\ref{fig:vary_first} and \ref{fig:vary_second}.

\section{Conclusion and Future Outlook}
\label{sec:conclusion}

Inside a core-collapse supernova, the neutrino density is so high that the neutrino matter potential receives a significant contribution from self-interaction leading to a phenomenon called collective oscillation. This alters the flavor contents of the neutrino flux emitted from the CCSN. In this article, we have studied the effects of spacetime geometry on neutrino oscillation inside a core-collapse supernova. We have assumed that outside the PNS the fermion gas is dense enough that the effective four-fermion torsional interaction, which applies to all fermions --- neutrinos, leptons, and quarks --- cannot be ignored. However, the spacetime curvature is small outside the PNS, enough that we can ignore the torsion-free part of the spin connection.

Since it is a current-current interaction of fermions in the mass basis, we cannot reduce its effect to that of a small mixing angle in the neutrino sector as is done when only weak interactions are considered. We have shown that, in the presence of neutrinos at uniform density, the survival probability $P_{\nu_e \to \nu_e}$ is periodic but the position of its first dip depends on the values of the geometrical coupling constants. We have shown that this coupling can break stability in flavor space if the mass spectrum of neutrino follows normal ordering.

We have also investigated the effects of this interaction in a realistic supernova where the density of neutrinos and antineutrinos fall off with distance from the core. We have shown that in the presence of uniform background density of fermions, the survival probability can change significantly depending on the coupling constants.

With several Mton detectors under operation or construction, we are entering the era of precision neutrino astronomy. CCSN can be a unique and interesting place to probe into neutrino flavor dynamics due to forces beyond the standard model. We see from this article that the contribution of the spacetime geometry significantly modifies the oscillation spectra and hence the flavor composition of the flux coming from a CCSN. This can be a promising place to study the bounds of the spin-torsion parameters.

{It is beyond the scope of this paper to consider oscillations into $\nu_\tau$ and $\bar{\nu}_\tau$\,, or the effect of any angular dependence of the background fermion density. } Only a full simulation of the flavor evolution {which include these } effects can give the precise flavor content of the neutrino flux coming to earth from a CCSN. 

\section{Acknowledgment}
\label{sec:acknowledgment}

I.G. would like to thank Riya Barick, Manibrata Sen, Anirban Das, and Abinash Medhi for the fruitful discussions. Part of the calculations were carried out with High-Performance Computing Facilities at SNBNCBS. The authors thank N.~Soker for making them aware of an alternative  to the stalled explosion, and the anonymous referee for suggesting several references.

\appendix
\section{Hamiltonian of neutrino-neutrino interaction}
\label{app:hamiltonian}
We derive the effective Hamiltonian due to neutrino self-interaction.
\subsection{Geometrical interaction}
\label{subsec:geometric}
\begin{figure}[hbtp]
     \centering
     \begin{subfigure}[b]{0.25\columnwidth}
         \centering
         \includegraphics[width=\textwidth]{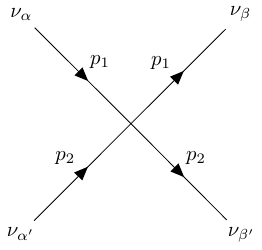}
         \caption{neutrino-neutrino scattering}
         \label{fig:nunu_fig}
     \end{subfigure} \qquad 
     \begin{subfigure}[b]{0.25\columnwidth}
         \centering
         \includegraphics[width=\textwidth]{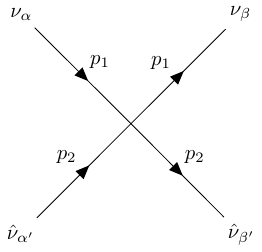}
         \caption{neutrino-antineutrino scattering}
         \label{fig:nunubar_fig}
     \end{subfigure}
        \caption{The two Feynman diagrams contributing to the neutrino self-interactions. $\hat{\nu}$ represents an antineutrino.}
        \label{fig:feynman_graphs}
\end{figure}

 We start from the four-fermion self-interaction Hamiltonian density of neutrinos corresponding to the Lagrangian density of the Eq.~\eqref{eq:spin-torsion_interaction},
\begin{align}
\Ha_{\operatorname{self}}=\frac{1}{4\sqrt{2}}\sum_{ij}\lambda_i\bar{\nu}_{i}\gamma^{\mu}(1-\gamma_5)\nu_i \lambda_j\bar{\nu}_{j}\gamma_{\mu}(1-\gamma_5)\nu_j \,.\label{eq:spin_torsion_interaction-Hamiltonian}
\end{align}
This Hamiltonian will contribute two terms in the self-interaction sector of the neutrino flavor evolution equation. These two terms can be expressed in terms of two Feynman diagrams shown in Fig.~\ref{fig:feynman_graphs}.
\subsubsection{Neutrino-neutrino potential} 

The matrix element corresponding to the Feynman diagram Fig.~\ref{fig:nunu_fig} is
\begin{align}
[H^{\nu\nu}_{\alpha \beta}]_{\alpha ^{\prime} \beta ^{\prime}}&=\int d^3x \frac{1}{4}\sum_{s_1,s_2} \bra{\nu_{\beta}(p_{1},s_{1})\nu_{\beta\p}(p_2,s_2)}\Ha_{\operatorname{self}}\ket{\nu_{\alpha}(p_{1},s_{1})\nu_{\alpha\p}(p_2,s_2)} \nn \\
&=\int d^3x \frac{1}{4}\sum_{s_1,s_2}\sum_{i_1,i_2}\sum_{j_1,j_2}\frac{1}{4\sqrt{2}}\sum_{a,b}V_{ab}^{i_1,i_2,j_1,j_2}U_{\beta j_1}U_{\beta^{\prime}  j_2}U^*_{\alpha i_1}U^*_{\alpha^{\prime} i_2} \label{eq:element_of_hamiltonian}
,\end{align}
where we have defined 
\begin{align}
V_{ab}^{i_1,i_2,j_1,j_2}=&\braket{\nu_{j_1}(p_1,s_1)\nu_{j_2}(p_2,s_2)|\lambda_{a}\bar{\nu}_{a}\gamma^{\mu}(1-\gamma_5)\nu_{a}\lambda_{b}\bar{\nu}_{b}\gamma_{\mu}(1-\gamma_5)\nu_{b}|\nu_{i_1}(p_1,s_1)\nu_{i_2}(p_2,s_2)} \nn \\
=&\int \frac{d^3p_{a} d^3p_{b} d^3k_{a} d^3k_{b}}{\sqrt{(2E_{p_{a}}V) (2E_{p_{b}}V) (2E_{k_{a}}V) (2E_{k_{b}}V)}} \nn \\
&\,\sum_{s_{a}s_{b}}\sum_{r_{a}r_{b}}\lambda_{a}\bar{u}_{a}(p_{a},s_{a})\gamma^{\mu}(1-\gamma_5)u_{a}(k_{a},r_{a})\lambda_{b}\bar{u}_{b}(p_{b},s_{b})\gamma_{\mu}(1-\gamma_5)u_{b}(k_{b},r_{b}) \nn \\
&\; e^{i(p_{a}-k_{a}+p_{b}-k_{b})\cdot x} \Big(\delta_{j_1a}\delta^3(p_1-p_{a})\delta_{s_1s_{a}}\delta_{j_2b}\delta^3(p_2-p_{b})\delta_{s_2s_{b}}- \nn \\
&\delta_{j_1b}\delta^3(p_1-p_{b})\delta_{s_1s_{b}}\delta_{j_2a}\delta^3(p_2-p_{a})\delta_{s_2s_{a}}\Big)\Big(\delta_{i_1a}\delta^3(p_1-k_{a})\delta_{s_1r_{a}}\delta_{i_2b}\delta^3(p_2-k_{b})\delta_{s_2r_{b}} \nn \\
&-\delta_{i_1b}\delta^3(p_1-k_{b})\delta_{s_1r_{b}}\delta_{i_2a}\delta^3(p_2-k_{a})\delta_{s_2r_{a}}\Big) \label{eq:hamiltonian_element_in_mass_space}
.\end{align}

We can evaluate the sums in Eq.~\eqref{eq:hamiltonian_element_in_mass_space} to write
\begin{align}
[H^{\nu\nu}_{\alpha\beta}]_{\alpha\p\beta\p}=&\sum_{s_1,s_2}\sum_{a,b}\frac{1}{4\sqrt{2}}\frac{1}{16E_1E_2V} \Big[-\lambda_{a}\bar{u}_{a}(p_2,s_2)\gamma^{\mu}(1-\gamma_5)u_{a}(p_1,s_1)\lambda_{b}\bar{u}_{b}(p_{1},s_1)\gamma_{\mu}(1-\gamma_5)u_{b}(p_2,s_2)2U_{\beta b}U_{\beta^{\prime}a}U^{*}_{\alpha a}U^{*}_{\alpha^{\prime}b} \nn \\
& \qquad + \lambda_{a}\bar{u}_{a}(p_1,s_1)\gamma^{\mu}(1-\gamma_5)u_{a}(p_1,s_1)\lambda_{b}\bar{u}_{b}(p_2,s_2)\gamma_{\mu}(1-\gamma_5)u_{b}(p_2,s_2)2U_{\beta a}U_{\beta^{\prime}b}U^{*}_{\alpha a}U^{*}_{\alpha^{\prime}b}\Big]
\label{eq:hamiltonian_in_terms_of_mass_basis_summed_over}
.\end{align}
The first part of the $[H^{\nu\nu}_{\alpha\beta}]_{\alpha\p\beta\p}$ can be rearranged by a Fierz transformation which changes the sign~\cite{Nieves:2004gfi}. We can now use the Gordon identity and fermion completeness relations to find
\begin{align}
[H^{\nu\nu}_{\alpha\beta}]_{\alpha^{\prime}\beta^{\prime}}&=\frac{\sqrt{2}}{4}\frac{1}{16E_1E_2V}16p_1^{\mu}p_{2\mu}
\left[\tilde{\Lambda}_{\alpha\beta}\tL_{\alpha\p\beta\p}+\tL_{\alpha\p\beta}\tL_{\alpha\beta\p}\right] \notag \\
&=\frac{\sqrt{2}}{4}\frac{1}{V}(1-\hat{p}_1\cdot \hat{p}_2) \left[\tilde{\Lambda}_{\alpha\beta}\tL_{\alpha\p\beta\p}+\tL_{\alpha\p\beta}\tL_{\alpha\beta\p}\right]\,, \label{eq:neutrino_neutrino_torsion_hamiltonian}
\end{align}
where $\tL=U^* \Lambda U^T$ and $\Lambda$ is a diagonal matrix whose entries are the coupling constants $\lambda_i$. In the case of the two neutrino family paradigm, {which we are using,} $\Lambda=\text{diag}(\lambda_1, \lambda_2)$.
%
\subsubsection{Neutrino-antineutrino potential} 
To derive this, we first note that the mixing matrix for antineutrinos is the complex conjugate of the mixing matrix for neutrinos. Then the matrix element corresponding to the Feynman diagram Fig.~\ref{fig:nunubar_fig} is
\begin{align}
[H^{\nu\bar{\nu}}_{\alpha\beta}]_{\alpha\p\beta\p}&=\int d^3x \frac{1}{4}\sum_{s_1,s_2} \bra{\nu_{\beta}(p_{1},s_{1})\hat{\nu}_{\beta\p}(p_2,s_2)}\Ha_{\operatorname{self}}\ket{\nu_{\alpha}(p_{1},s_{1})\hat{\nu}_{\alpha\p}(p_2,s_2)} \nn \\
&=\int d^3x \frac{1}{4}\sum_{s_1,s_2}\sum_{ab}\sum_{i_1,i_2}\sum_{j_1,j_2}\frac{1}{4\sqrt{2}}\tilde{V}^{i_1i_2j_1j_2}_{ab}U_{\beta j_1}U^{*}_{\beta\p j_2}U^{*}_{\alpha i_1}U_{\alpha\p i_2} \label{eq:element_of_hamiltonian1}
.\end{align}
Denoting antineutrinos as $\hat{\nu}$, we can write
\begin{align}
\tilde{V}_{ab}^{i_1i_2j_1j_2}&=\bra{\nu_{j_1}(p_1,s_1)\hat{\nu}_{j_2}(p_2,s_2)}\bar{\nu}_{a}\gamma^{\mu}\lambda_{a}(1-\gamma^5)\nu_{a}\bar{\nu}_{b}\gamma_{\mu}\lambda_{b}(1-\gamma^5)\nu_{b}\ket{\nu_{i_1}(p_1,s_1)\hat{\nu}_{i_2}(p_2,s_2)}
.\end{align}
Following the same procedure as before we have
\begin{align}
[H^{\nu\bar{\nu}}_{\alpha\beta}]_{\alpha\p\beta\p}=&\sum_{s_1,s_2}\sum_{a,b}\frac{1}{4\sqrt{2}}\frac{1}{16V}\Big(\lambda_{a}\bar{u}_{a}(p_1,s_1)\gamma^{\mu}(1-\gamma^5)v_{a}(p_2,s_2)\lambda_{b}\bar{v}_{b}(p_2,s_2)\gamma_{\mu}(1-\gamma^5)u_{b}(p_1,s_1)2U_{\beta a}U^{*}_{\beta\p a}U^{*}_{\alpha b}U_{\alpha\p b} \nn \\
&-\lambda_{a}\bar{u}_{a}(p_1,s_1)\gamma^{\mu}(1-\gamma_5)v_{a}(p_1,s_1)\lambda_{b}\bar{v}_{b}(p_2,s_2)\gamma_{\mu}(1-\gamma^5)u_{b}(p_2,s_2)2U_{\beta a}U^{*}_{\beta\p b}U^{*}_{\alpha a}U_{\alpha\p b}\Big)\,.\end{align}
We apply Fierz transformation to the first term and use the completeness relations of the spinors to get~\\
\begin{align}
[H^{\nu\nu}_{\alpha\beta}]_{\alpha\p\beta\p}=&\frac{\sqrt{2}}{4}\frac{1}{16E_1E_2V}16p_1^{\mu}p_{2\mu}(1-\hat{p}_1\cdot\hat{p}_2) \left(\tL_{\alpha\beta}\tL_{\beta\p\alpha\p}+\tL_{\alpha\alpha\p}\tL_{\beta\p\beta}\right) \notag \\
=&-\frac{\sqrt{2}}{4}\frac{1}{V}(1-\hat{p}_1\cdot\hat{p}_2)\left(\tL_{\alpha\beta}\tL_{\beta\p\alpha\p}+\tL_{\alpha\alpha\p}\tL_{\beta\p\beta}\right) \label{eq:elements_of_hamiltonian1}
.\end{align}

\subsubsection{Liouville-von Neumann equation}\label{density-matrix}

{The Hamiltonians $H^{\nu\nu}$ and $H^{\nu\bar{\nu}}$ that we calculated above act} on two particle states. Hence, to see the evolution of flavors of single-particle states {in a background of other particles,} we have to take a partial trace over two of its indices. The partial trace will be carried over one particle in the initial state and one in the final.

Following the framework described in~\cite{Pantaleone:1994ns, Dasgupta:2007ws}, we assume that the neutrinos are described by a local density matrix $\rho$ and the antineutrinos are described by $\bar{\rho}$. In our conventions all the states --- vacuum, single particle, or multiple particles, are dimensionless. The multiparticle density matrices are normalized to their corresponding phase space number density. Hence, $\operatorname{Tr}(\rho(p))=n(p,s,x)$, where $n(p,s,x)$ is the number density of neutrinos in phase space irrespective of flavor. This means $\int d^3x~d^3p~n(p,s,x) = N$, where $N$ is the total number of neutrinos irrespective of flavor in the whole position and momentum volume. Similarly, $\operatorname{Tr}(\rho(p))=\bar{n}(p,s,x)$, where $\bar{n}(p,s,x)$ is the number density of antineutrinos in phase space irrespective of flavor. We assume the neutrino violates parity maximally and hence the spin is not an independent variable. Also, in the flavor basis, the local density matrices are defined as

\begin{align}
\rho_{\beta\alpha}(p_1, s_1, x)&=C_{\alpha\beta}(p_1, s_1, x)\ket{\nu_{\alpha}(p_1,s_1)}\bra{\nu_{\beta}(p_1,s_1)},  \\
\bar{\rho}_{\alpha\beta}(p_1, s_1, x)&=\bar{C}_{\alpha\beta}(p_1, s_1, x)\ket{\hat{\nu}_{\alpha}(p_1,s_1)}\bra{\hat{\nu}_{\beta}(p_1,s_1)}
.\end{align}
We assume that the local density matrix can be expanded in the plane wave basis at each point in the phase space. We will denote the momentum of the states getting traced over by $p_2$. Finally, we integrate the momentum $p_2$ to cover all the neutrinos and antineutrinos. Thus the Hamiltonian $H^S_{\nu\nu}$ evolving the single neutrino states due to self-interaction due to the Hamiltonian density given by Eq. \eqref{eq:spin_torsion_interaction-Hamiltonian} is given by

\begin{align}
[H_{\nu\nu}^{S}]_{\alpha \beta}=&V \int d^3p_2 \operatorname{Tr}_B \biggl(\int d^3x \Ha_{\operatorname{self}} \ket{\nu_{\alpha}(p_1,s_1)}\bra{\nu_{\beta}(p_1,s_1)} \otimes \rho_B (p_2,s_2, x)\biggr) \notag \\
&+V \int d^3p_2 \operatorname{Tr}_B \biggl(\int d^3x \Ha_{\operatorname{self}} \ket{\nu_{\alpha}(p_1,s_1)}\bra{\nu_{\beta}(p_1,s_1)} \otimes \bar{\rho}_{B}(p_2, s_2, x) \biggr) \notag \\
=&\int d^3p_2 \sum_{\alpha\p \beta\p} \frac{\sqrt{2}}{4}(1-\hat{p}_1\cdot\hat{p}_2) \left(\tL_{\alpha\beta}\tL_{\alpha\p\beta\p}+\tL_{\alpha\beta\p}\tL_{\alpha\p\beta}\right)\rho_{\beta\p\alpha\p}(p_2) \nn \\
&-\int d^3p_2 \sum_{\alpha\p \beta\p} \frac{\sqrt{2}}{4}(1-\hat{p}_1\cdot\hat{p}_2)\left(\tL_{\alpha\beta}\tL_{\beta\p\alpha\p}+\tL_{\alpha\alpha\p}\tL_{\beta\p\beta}\right)\bar{\rho}_{\alpha\p\beta\p}(p_2) \nn \\
=&\int d^3p_2 (1-\hat{p}_1\cdot\hat{p}_2)\frac{\sqrt{2}}{4}\left(\rm{Tr}\left[\tL(\rho-\bar{\rho})\right]\tL_{\alpha\beta}+(\tL(\rho-\bar{\rho})\tL)_{\alpha\beta}\right)\,. \label{eq:self_torsion}
\end{align}
We assume that the density matrix depends only on the magnitude of the spatial components of $p_2$, so
the $\hat{p}_1\cdot\hat{p}_2$ term drops out after integration over $p_2$. Now, in this text, we will assume all the neutrinos have only one single energy. Then the integration only leaves out one mode. The contribution from Eq.~\eqref{eq:self_torsion} is now
\begin{align}
    (H_{\nu\nu}^{S})_{\alpha\beta}=\frac{\sqrt{2}}{4}\left(\rm{Tr}\left[\tL(\rho-\bar{\rho})\right]\tL_{\alpha\beta}+(\tL(\rho-\bar{\rho})\tL)_{\alpha\beta}\right) \label{eq:self_torsion_single_mode}
.\end{align}

\subsection{Weak interaction}
\label{subsec:weak}

In the standard model, the neutral vector boson part of the Hamiltonian will contribute to the neutrino self-interaction. The neutral current exchange Hamiltonian can be written at the tree level as 

\begin{align}
\Ha&=\frac{G_F}{\sqrt{2}}\sum_{\alpha,\beta}\bar{\nu}_{\alpha}\gamma^{\mu}(1-\gamma_5)\nu_{\alpha}\bar{\nu}_{\beta}\gamma_{\mu}(1-\gamma_5)\nu_{\beta} \nn \\
&=\frac{G_F}{\sqrt{2}}\sum_{\alpha,\beta}\sum_{i_1,i_2}\sum_{j_1,j_2}U_{\alpha i_1}U^*_{\alpha i_2}\bar{\nu}_{i_1}\gamma^{\mu}(1-\gamma_5)\nu_{i_2}U_{\beta j_1}U^{*}_{\beta j_2}\bar{\nu}_{j_1}\gamma_{\mu}(1-\gamma_5)\nu_{j_2} \nn \\
&=\frac{G_F}{\sqrt{2}}\sum_{i,j}\bar{\nu}_{i}\gamma^{\mu}(1-\gamma_5)\nu_{i}\bar{\nu}_{j}\gamma_{\mu}(1-\gamma_5)\nu_{j}\,.\label{eq:flavor_to_mass_hamiltonian}
\end{align}

Comparing Eq.~\eqref{eq:flavor_to_mass_hamiltonian} and  \eqref{eq:spin_torsion_interaction-Hamiltonian} we see that the neutrino self-interaction Hamiltonian coming from weak interactions is similar to the geometric self-interaction but with the universal coupling $2\sqrt{G_F}$ taking the place of all the $\lambda_i$'s. From this, we see that the equation governing flavor evolution due to self-interaction mediated by the weak interaction would be
\begin{align}
[H_{\nu\nu}^{W}]_{\alpha \beta}=&  \frac{1}{2} \sum_{s_2} \int d^3p_2 \sum_{\alpha\p \beta\p} \sqrt{2}G_F(1-\hat{p}_1\cdot\hat{p}_2)(\delta_{\alpha\beta}\delta_{\alpha\p\beta\p}+\delta_{\alpha\beta\p}\delta_{\alpha\p\beta})\rho_{\beta\p\alpha\p} \nn \\
&-\frac{1}{2} \sum_{s_2} \int d^3p_2 \sum_{\alpha\p \beta\p} \sqrt{2}G_F(1-\hat{p}_1\cdot\hat{p}_2)(\delta_{\alpha\beta}\delta_{\beta\p\alpha\p}+\delta_{\alpha\alpha\p}\delta_{\beta\p\beta})\bar{\rho}_{\beta\p\alpha\p} \nn \\
=& \frac{1}{2} \sum_{s_2} \int d^3p_2 \sqrt{2}G_F(1-\hat{p}_1\cdot\hat{p}_2) (\rho - \bar{\rho})_{\alpha\beta}\,. \label{eq:self_weak}
\end{align}
It is to be noted that we have suppressed any term proportional to identity. {A similar calculation to that done in Appendix~{\ref{density-matrix}}} above reduces the expression Eq.~\eqref{eq:self_weak} to
\begin{align}
    H_{\nu\nu}^W=\sqrt{2}G_F(\rho-\bar{\rho})\,, \label{eq:self_weak_single_mode}
\end{align}
which agrees with the result derived in~\cite{Neto:2021hhl}.

\bibliography{references}

\begin{thebibliography}{65}%
\makeatletter
\providecommand \@ifxundefined [1]{%
 \@ifx{#1\undefined}
}%
\providecommand \@ifnum [1]{%
 \ifnum #1\expandafter \@firstoftwo
 \else \expandafter \@secondoftwo
 \fi
}%
\providecommand \@ifx [1]{%
 \ifx #1\expandafter \@firstoftwo
 \else \expandafter \@secondoftwo
 \fi
}%
\providecommand \natexlab [1]{#1}%
\providecommand \enquote  [1]{``#1''}%
\providecommand \bibnamefont  [1]{#1}%
\providecommand \bibfnamefont [1]{#1}%
\providecommand \citenamefont [1]{#1}%
\providecommand \href@noop [0]{\@secondoftwo}%
\providecommand \href [0]{\begingroup \@sanitize@url \@href}%
\providecommand \@href[1]{\@@startlink{#1}\@@href}%
\providecommand \@@href[1]{\endgroup#1\@@endlink}%
\providecommand \@sanitize@url [0]{\catcode `\\12\catcode `\$12\catcode `\&12\catcode `\#12\catcode `\^12\catcode `\_12\catcode `\%12\relax}%
\providecommand \@@startlink[1]{}%
\providecommand \@@endlink[0]{}%
\providecommand \url  [0]{\begingroup\@sanitize@url \@url }%
\providecommand \@url [1]{\endgroup\@href {#1}{\urlprefix }}%
\providecommand \urlprefix  [0]{URL }%
\providecommand \Eprint [0]{\href }%
\providecommand \doibase [0]{http://dx.doi.org/}%
\providecommand \selectlanguage [0]{\@gobble}%
\providecommand \bibinfo  [0]{\@secondoftwo}%
\providecommand \bibfield  [0]{\@secondoftwo}%
\providecommand \translation [1]{[#1]}%
\providecommand \BibitemOpen [0]{}%
\providecommand \bibitemStop [0]{}%
\providecommand \bibitemNoStop [0]{.\EOS\space}%
\providecommand \EOS [0]{\spacefactor3000\relax}%
\providecommand \BibitemShut  [1]{\csname bibitem#1\endcsname}%
\let\auto@bib@innerbib\@empty
\bibitem [{\citenamefont {Colgate}\ and\ \citenamefont {White}(1966)}]{Colgate:1966ax}%
  \BibitemOpen
  \bibfield  {author} {\bibinfo {author} {\bibfnamefont {S.~A.}\ \bibnamefont {Colgate}}\ and\ \bibinfo {author} {\bibfnamefont {R.~H.}\ \bibnamefont {White}},\ }\href {\doibase 10.1086/148549} {\bibfield  {journal} {\bibinfo  {journal} {Astrophys. J.}\ }\textbf {\bibinfo {volume} {143}},\ \bibinfo {pages} {626} (\bibinfo {year} {1966})}\BibitemShut {NoStop}%
\bibitem [{\citenamefont {Herant}\ \emph {et~al.}(1994)\citenamefont {Herant}, \citenamefont {Benz}, \citenamefont {Hix}, \citenamefont {Fryer},\ and\ \citenamefont {Colgate}}]{Herant:1994dd}%
  \BibitemOpen
  \bibfield  {author} {\bibinfo {author} {\bibfnamefont {M.}~\bibnamefont {Herant}}, \bibinfo {author} {\bibfnamefont {W.}~\bibnamefont {Benz}}, \bibinfo {author} {\bibfnamefont {W.~R.}\ \bibnamefont {Hix}}, \bibinfo {author} {\bibfnamefont {C.~L.}\ \bibnamefont {Fryer}}, \ and\ \bibinfo {author} {\bibfnamefont {S.~A.}\ \bibnamefont {Colgate}},\ }\href {\doibase 10.1086/174817} {\bibfield  {journal} {\bibinfo  {journal} {Astrophys. J.}\ }\textbf {\bibinfo {volume} {435}},\ \bibinfo {pages} {339} (\bibinfo {year} {1994})},\ \Eprint {http://arxiv.org/abs/astro-ph/9404024} {arXiv:astro-ph/9404024} \BibitemShut {NoStop}%
\bibitem [{\citenamefont {Janka}\ \emph {et~al.}(2007)\citenamefont {Janka}, \citenamefont {Langanke}, \citenamefont {Marek}, \citenamefont {Martinez-Pinedo},\ and\ \citenamefont {Mueller}}]{Janka:2006fh}%
  \BibitemOpen
  \bibfield  {author} {\bibinfo {author} {\bibfnamefont {H.-T.}\ \bibnamefont {Janka}}, \bibinfo {author} {\bibfnamefont {K.}~\bibnamefont {Langanke}}, \bibinfo {author} {\bibfnamefont {A.}~\bibnamefont {Marek}}, \bibinfo {author} {\bibfnamefont {G.}~\bibnamefont {Martinez-Pinedo}}, \ and\ \bibinfo {author} {\bibfnamefont {B.}~\bibnamefont {Mueller}},\ }\href {\doibase 10.1016/j.physrep.2007.02.002} {\bibfield  {journal} {\bibinfo  {journal} {Phys. Rept.}\ }\textbf {\bibinfo {volume} {442}},\ \bibinfo {pages} {38} (\bibinfo {year} {2007})},\ \Eprint {http://arxiv.org/abs/astro-ph/0612072} {arXiv:astro-ph/0612072} \BibitemShut {NoStop}%
\bibitem [{\citenamefont {Chakraborty}\ \emph {et~al.}(2011)\citenamefont {Chakraborty}, \citenamefont {Fischer}, \citenamefont {Mirizzi}, \citenamefont {Saviano},\ and\ \citenamefont {Tomas}}]{Chakraborty:2011gd}%
  \BibitemOpen
  \bibfield  {author} {\bibinfo {author} {\bibfnamefont {S.}~\bibnamefont {Chakraborty}}, \bibinfo {author} {\bibfnamefont {T.}~\bibnamefont {Fischer}}, \bibinfo {author} {\bibfnamefont {A.}~\bibnamefont {Mirizzi}}, \bibinfo {author} {\bibfnamefont {N.}~\bibnamefont {Saviano}}, \ and\ \bibinfo {author} {\bibfnamefont {R.}~\bibnamefont {Tomas}},\ }\href {\doibase 10.1103/PhysRevD.84.025002} {\bibfield  {journal} {\bibinfo  {journal} {Phys. Rev. D}\ }\textbf {\bibinfo {volume} {84}},\ \bibinfo {pages} {025002} (\bibinfo {year} {2011})},\ \Eprint {http://arxiv.org/abs/1105.1130} {arXiv:1105.1130 [hep-ph]} \BibitemShut {NoStop}%
\bibitem [{\citenamefont {Janka}(2012)}]{Janka:2012wk}%
  \BibitemOpen
  \bibfield  {author} {\bibinfo {author} {\bibfnamefont {H.-T.}\ \bibnamefont {Janka}},\ }\href {\doibase 10.1146/annurev-nucl-102711-094901} {\bibfield  {journal} {\bibinfo  {journal} {Ann. Rev. Nucl. Part. Sci.}\ }\textbf {\bibinfo {volume} {62}},\ \bibinfo {pages} {407} (\bibinfo {year} {2012})},\ \Eprint {http://arxiv.org/abs/1206.2503} {arXiv:1206.2503 [astro-ph.SR]} \BibitemShut {NoStop}%
\bibitem [{\citenamefont {Janka}\ \emph {et~al.}(2012)\citenamefont {Janka}, \citenamefont {Hanke}, \citenamefont {Huedepohl}, \citenamefont {Marek}, \citenamefont {Mueller},\ and\ \citenamefont {Obergaulinger}}]{Janka:2012sb}%
  \BibitemOpen
  \bibfield  {author} {\bibinfo {author} {\bibfnamefont {H.~T.}\ \bibnamefont {Janka}}, \bibinfo {author} {\bibfnamefont {F.}~\bibnamefont {Hanke}}, \bibinfo {author} {\bibfnamefont {L.}~\bibnamefont {Huedepohl}}, \bibinfo {author} {\bibfnamefont {A.}~\bibnamefont {Marek}}, \bibinfo {author} {\bibfnamefont {B.}~\bibnamefont {Mueller}}, \ and\ \bibinfo {author} {\bibfnamefont {M.}~\bibnamefont {Obergaulinger}},\ }\href {\doibase 10.1093/ptep/pts067} {\bibfield  {journal} {\bibinfo  {journal} {PTEP}\ }\textbf {\bibinfo {volume} {2012}},\ \bibinfo {pages} {01A309} (\bibinfo {year} {2012})},\ \Eprint {http://arxiv.org/abs/1211.1378} {arXiv:1211.1378 [astro-ph.SR]} \BibitemShut {NoStop}%
\bibitem [{\citenamefont {Volpe}(2024)}]{Volpe:2023met}%
  \BibitemOpen
  \bibfield  {author} {\bibinfo {author} {\bibfnamefont {M.~C.}\ \bibnamefont {Volpe}},\ }\href {\doibase 10.1103/RevModPhys.96.025004} {\bibfield  {journal} {\bibinfo  {journal} {Rev. Mod. Phys.}\ }\textbf {\bibinfo {volume} {96}},\ \bibinfo {pages} {025004} (\bibinfo {year} {2024})},\ \Eprint {http://arxiv.org/abs/2301.11814} {arXiv:2301.11814 [hep-ph]} \BibitemShut {NoStop}%
\bibitem [{\citenamefont {Johns}\ \emph {et~al.}(2025)\citenamefont {Johns}, \citenamefont {Richers},\ and\ \citenamefont {Wu}}]{Johns:2025mlm}%
  \BibitemOpen
  \bibfield  {author} {\bibinfo {author} {\bibfnamefont {L.}~\bibnamefont {Johns}}, \bibinfo {author} {\bibfnamefont {S.}~\bibnamefont {Richers}}, \ and\ \bibinfo {author} {\bibfnamefont {M.-R.}\ \bibnamefont {Wu}},\ }\href {\doibase 10.1146/annurev-nucl-121423-100853} {\  (\bibinfo {year} {2025}),\ 10.1146/annurev-nucl-121423-100853},\ \Eprint {http://arxiv.org/abs/2503.05959} {arXiv:2503.05959 [astro-ph.HE]} \BibitemShut {NoStop}%
\bibitem [{\citenamefont {Qian}\ \emph {et~al.}(1993)\citenamefont {Qian}, \citenamefont {Fuller}, \citenamefont {Mathews}, \citenamefont {Mayle}, \citenamefont {Wilson},\ and\ \citenamefont {Woosley}}]{Qian:1993dg}%
  \BibitemOpen
  \bibfield  {author} {\bibinfo {author} {\bibfnamefont {Y.-Z.}\ \bibnamefont {Qian}}, \bibinfo {author} {\bibfnamefont {G.~M.}\ \bibnamefont {Fuller}}, \bibinfo {author} {\bibfnamefont {G.~J.}\ \bibnamefont {Mathews}}, \bibinfo {author} {\bibfnamefont {R.}~\bibnamefont {Mayle}}, \bibinfo {author} {\bibfnamefont {J.~R.}\ \bibnamefont {Wilson}}, \ and\ \bibinfo {author} {\bibfnamefont {S.~E.}\ \bibnamefont {Woosley}},\ }\href {\doibase 10.1103/PhysRevLett.71.1965} {\bibfield  {journal} {\bibinfo  {journal} {Phys. Rev. Lett.}\ }\textbf {\bibinfo {volume} {71}},\ \bibinfo {pages} {1965} (\bibinfo {year} {1993})}\BibitemShut {NoStop}%
\bibitem [{\citenamefont {Chakraborty}\ \emph {et~al.}(2010)\citenamefont {Chakraborty}, \citenamefont {Choubey}, \citenamefont {Goswami},\ and\ \citenamefont {Kar}}]{Chakraborty:2009ej}%
  \BibitemOpen
  \bibfield  {author} {\bibinfo {author} {\bibfnamefont {S.}~\bibnamefont {Chakraborty}}, \bibinfo {author} {\bibfnamefont {S.}~\bibnamefont {Choubey}}, \bibinfo {author} {\bibfnamefont {S.}~\bibnamefont {Goswami}}, \ and\ \bibinfo {author} {\bibfnamefont {K.}~\bibnamefont {Kar}},\ }\href {\doibase 10.1088/1475-7516/2010/06/007} {\bibfield  {journal} {\bibinfo  {journal} {JCAP}\ }\textbf {\bibinfo {volume} {06}},\ \bibinfo {pages} {007} (\bibinfo {year} {2010})},\ \Eprint {http://arxiv.org/abs/0911.1218} {arXiv:0911.1218 [hep-ph]} \BibitemShut {NoStop}%
\bibitem [{\citenamefont {Duan}\ \emph {et~al.}(2011)\citenamefont {Duan}, \citenamefont {Friedland}, \citenamefont {McLaughlin},\ and\ \citenamefont {Surman}}]{Duan:2010af}%
  \BibitemOpen
  \bibfield  {author} {\bibinfo {author} {\bibfnamefont {H.}~\bibnamefont {Duan}}, \bibinfo {author} {\bibfnamefont {A.}~\bibnamefont {Friedland}}, \bibinfo {author} {\bibfnamefont {G.}~\bibnamefont {McLaughlin}}, \ and\ \bibinfo {author} {\bibfnamefont {R.}~\bibnamefont {Surman}},\ }\href {\doibase 10.1088/0954-3899/38/3/035201} {\bibfield  {journal} {\bibinfo  {journal} {J. Phys. G}\ }\textbf {\bibinfo {volume} {38}},\ \bibinfo {pages} {035201} (\bibinfo {year} {2011})},\ \Eprint {http://arxiv.org/abs/1012.0532} {arXiv:1012.0532 [astro-ph.SR]} \BibitemShut {NoStop}%
\bibitem [{\citenamefont {Mart\'\i{}nez-Pinedo}\ \emph {et~al.}(2014)\citenamefont {Mart\'\i{}nez-Pinedo}, \citenamefont {Fischer},\ and\ \citenamefont {Huther}}]{Martinez-Pinedo:2013jna}%
  \BibitemOpen
  \bibfield  {author} {\bibinfo {author} {\bibfnamefont {G.}~\bibnamefont {Mart\'\i{}nez-Pinedo}}, \bibinfo {author} {\bibfnamefont {T.}~\bibnamefont {Fischer}}, \ and\ \bibinfo {author} {\bibfnamefont {L.}~\bibnamefont {Huther}},\ }\href {\doibase 10.1088/0954-3899/41/4/044008} {\bibfield  {journal} {\bibinfo  {journal} {J. Phys. G}\ }\textbf {\bibinfo {volume} {41}},\ \bibinfo {pages} {044008} (\bibinfo {year} {2014})},\ \Eprint {http://arxiv.org/abs/1309.5477} {arXiv:1309.5477 [astro-ph.HE]} \BibitemShut {NoStop}%
\bibitem [{\citenamefont {Wu}\ \emph {et~al.}(2014)\citenamefont {Wu}, \citenamefont {Fischer}, \citenamefont {Huther}, \citenamefont {Mart\'\i{}nez-Pinedo},\ and\ \citenamefont {Qian}}]{Wu:2013gxa}%
  \BibitemOpen
  \bibfield  {author} {\bibinfo {author} {\bibfnamefont {M.-R.}\ \bibnamefont {Wu}}, \bibinfo {author} {\bibfnamefont {T.}~\bibnamefont {Fischer}}, \bibinfo {author} {\bibfnamefont {L.}~\bibnamefont {Huther}}, \bibinfo {author} {\bibfnamefont {G.}~\bibnamefont {Mart\'\i{}nez-Pinedo}}, \ and\ \bibinfo {author} {\bibfnamefont {Y.-Z.}\ \bibnamefont {Qian}},\ }\href {\doibase 10.1103/PhysRevD.89.061303} {\bibfield  {journal} {\bibinfo  {journal} {Phys. Rev. D}\ }\textbf {\bibinfo {volume} {89}},\ \bibinfo {pages} {061303} (\bibinfo {year} {2014})},\ \Eprint {http://arxiv.org/abs/1305.2382} {arXiv:1305.2382 [astro-ph.HE]} \BibitemShut {NoStop}%
\bibitem [{\citenamefont {Wu}\ \emph {et~al.}(2015)\citenamefont {Wu}, \citenamefont {Qian}, \citenamefont {Martinez-Pinedo}, \citenamefont {Fischer},\ and\ \citenamefont {Huther}}]{Wu:2014kaa}%
  \BibitemOpen
  \bibfield  {author} {\bibinfo {author} {\bibfnamefont {M.-R.}\ \bibnamefont {Wu}}, \bibinfo {author} {\bibfnamefont {Y.-Z.}\ \bibnamefont {Qian}}, \bibinfo {author} {\bibfnamefont {G.}~\bibnamefont {Martinez-Pinedo}}, \bibinfo {author} {\bibfnamefont {T.}~\bibnamefont {Fischer}}, \ and\ \bibinfo {author} {\bibfnamefont {L.}~\bibnamefont {Huther}},\ }\href {\doibase 10.1103/PhysRevD.91.065016} {\bibfield  {journal} {\bibinfo  {journal} {Phys. Rev. D}\ }\textbf {\bibinfo {volume} {91}},\ \bibinfo {pages} {065016} (\bibinfo {year} {2015})},\ \Eprint {http://arxiv.org/abs/1412.8587} {arXiv:1412.8587 [astro-ph.HE]} \BibitemShut {NoStop}%
\bibitem [{\citenamefont {Sasaki}\ \emph {et~al.}(2017)\citenamefont {Sasaki}, \citenamefont {Kajino}, \citenamefont {Takiwaki}, \citenamefont {Hayakawa}, \citenamefont {Balantekin},\ and\ \citenamefont {Pehlivan}}]{Sasaki:2017jry}%
  \BibitemOpen
  \bibfield  {author} {\bibinfo {author} {\bibfnamefont {H.}~\bibnamefont {Sasaki}}, \bibinfo {author} {\bibfnamefont {T.}~\bibnamefont {Kajino}}, \bibinfo {author} {\bibfnamefont {T.}~\bibnamefont {Takiwaki}}, \bibinfo {author} {\bibfnamefont {T.}~\bibnamefont {Hayakawa}}, \bibinfo {author} {\bibfnamefont {A.~B.}\ \bibnamefont {Balantekin}}, \ and\ \bibinfo {author} {\bibfnamefont {Y.}~\bibnamefont {Pehlivan}},\ }\href {\doibase 10.1103/PhysRevD.96.043013} {\bibfield  {journal} {\bibinfo  {journal} {Phys. Rev. D}\ }\textbf {\bibinfo {volume} {96}},\ \bibinfo {pages} {043013} (\bibinfo {year} {2017})},\ \Eprint {http://arxiv.org/abs/1707.09111} {arXiv:1707.09111 [astro-ph.HE]} \BibitemShut {NoStop}%
\bibitem [{\citenamefont {Xiong}\ \emph {et~al.}(2019)\citenamefont {Xiong}, \citenamefont {Wu},\ and\ \citenamefont {Qian}}]{Xiong:2019nvw}%
  \BibitemOpen
  \bibfield  {author} {\bibinfo {author} {\bibfnamefont {Z.}~\bibnamefont {Xiong}}, \bibinfo {author} {\bibfnamefont {M.-R.}\ \bibnamefont {Wu}}, \ and\ \bibinfo {author} {\bibfnamefont {Y.-Z.}\ \bibnamefont {Qian}},\ }\href {\doibase 10.3847/1538-4357/ab2870} {\bibfield  {journal} {\bibinfo  {journal} {Astrophys. J.}\ } (\bibinfo {year} {2019}),\ 10.3847/1538-4357/ab2870},\ \Eprint {http://arxiv.org/abs/1904.09371} {arXiv:1904.09371 [astro-ph.HE]} \BibitemShut {NoStop}%
\bibitem [{\citenamefont {Ko}\ \emph {et~al.}(2020)\citenamefont {Ko} \emph {et~al.}}]{Ko:2019xxm}%
  \BibitemOpen
  \bibfield  {author} {\bibinfo {author} {\bibfnamefont {H.}~\bibnamefont {Ko}} \emph {et~al.},\ }\href {\doibase 10.3847/2041-8213/ab775b} {\bibfield  {journal} {\bibinfo  {journal} {Astrophys. J. Lett.}\ }\textbf {\bibinfo {volume} {891}},\ \bibinfo {pages} {L24} (\bibinfo {year} {2020})},\ \Eprint {http://arxiv.org/abs/1903.02086} {arXiv:1903.02086 [astro-ph.HE]} \BibitemShut {NoStop}%
\bibitem [{\citenamefont {Xiong}\ \emph {et~al.}(2020)\citenamefont {Xiong}, \citenamefont {Sieverding}, \citenamefont {Sen},\ and\ \citenamefont {Qian}}]{Xiong:2020ntn}%
  \BibitemOpen
  \bibfield  {author} {\bibinfo {author} {\bibfnamefont {Z.}~\bibnamefont {Xiong}}, \bibinfo {author} {\bibfnamefont {A.}~\bibnamefont {Sieverding}}, \bibinfo {author} {\bibfnamefont {M.}~\bibnamefont {Sen}}, \ and\ \bibinfo {author} {\bibfnamefont {Y.-Z.}\ \bibnamefont {Qian}},\ }\href {\doibase 10.3847/1538-4357/abac5e} {\bibfield  {journal} {\bibinfo  {journal} {Astrophys. J.}\ }\textbf {\bibinfo {volume} {900}},\ \bibinfo {pages} {144} (\bibinfo {year} {2020})},\ \Eprint {http://arxiv.org/abs/2006.11414} {arXiv:2006.11414 [astro-ph.HE]} \BibitemShut {NoStop}%
\bibitem [{\citenamefont {George}\ \emph {et~al.}(2020)\citenamefont {George}, \citenamefont {Wu}, \citenamefont {Tamborra}, \citenamefont {Ardevol-Pulpillo},\ and\ \citenamefont {Janka}}]{George:2020veu}%
  \BibitemOpen
  \bibfield  {author} {\bibinfo {author} {\bibfnamefont {M.}~\bibnamefont {George}}, \bibinfo {author} {\bibfnamefont {M.-R.}\ \bibnamefont {Wu}}, \bibinfo {author} {\bibfnamefont {I.}~\bibnamefont {Tamborra}}, \bibinfo {author} {\bibfnamefont {R.}~\bibnamefont {Ardevol-Pulpillo}}, \ and\ \bibinfo {author} {\bibfnamefont {H.-T.}\ \bibnamefont {Janka}},\ }\href {\doibase 10.1103/PhysRevD.102.103015} {\bibfield  {journal} {\bibinfo  {journal} {Phys. Rev. D}\ }\textbf {\bibinfo {volume} {102}},\ \bibinfo {pages} {103015} (\bibinfo {year} {2020})},\ \Eprint {http://arxiv.org/abs/2009.04046} {arXiv:2009.04046 [astro-ph.HE]} \BibitemShut {NoStop}%
\bibitem [{\citenamefont {Mezzacappa}\ and\ \citenamefont {Messer}(1999)}]{MEZZACAPPA1999281}%
  \BibitemOpen
  \bibfield  {author} {\bibinfo {author} {\bibfnamefont {A.}~\bibnamefont {Mezzacappa}}\ and\ \bibinfo {author} {\bibfnamefont {O.}~\bibnamefont {Messer}},\ }\href {\doibase https://doi.org/10.1016/S0377-0427(99)00162-4} {\bibfield  {journal} {\bibinfo  {journal} {Journal of Computational and Applied Mathematics}\ }\textbf {\bibinfo {volume} {109}},\ \bibinfo {pages} {281} (\bibinfo {year} {1999})}\BibitemShut {NoStop}%
\bibitem [{\citenamefont {Janka}(2025)}]{Janka:2025tvf}%
  \BibitemOpen
  \bibfield  {author} {\bibinfo {author} {\bibfnamefont {H.~T.}\ \bibnamefont {Janka}},\ }\href@noop {} {\  (\bibinfo {year} {2025})},\ \Eprint {http://arxiv.org/abs/2502.14836} {arXiv:2502.14836 [astro-ph.HE]} \BibitemShut {NoStop}%
\bibitem [{\citenamefont {Betranhandy}\ and\ \citenamefont {O'Connor}(2022)}]{Betranhandy:2022bvr}%
  \BibitemOpen
  \bibfield  {author} {\bibinfo {author} {\bibfnamefont {A.}~\bibnamefont {Betranhandy}}\ and\ \bibinfo {author} {\bibfnamefont {E.}~\bibnamefont {O'Connor}},\ }\href {\doibase 10.1103/PhysRevD.106.063019} {\bibfield  {journal} {\bibinfo  {journal} {Phys. Rev. D}\ }\textbf {\bibinfo {volume} {106}},\ \bibinfo {pages} {063019} (\bibinfo {year} {2022})},\ \Eprint {http://arxiv.org/abs/2204.00503} {arXiv:2204.00503 [astro-ph.HE]} \BibitemShut {NoStop}%
\bibitem [{\citenamefont {Couch}(2017)}]{couch2017mechanism}%
  \BibitemOpen
  \bibfield  {author} {\bibinfo {author} {\bibfnamefont {S.~M.}\ \bibnamefont {Couch}},\ }\href@noop {} {\bibfield  {journal} {\bibinfo  {journal} {Philosophical Transactions of the Royal Society A: Mathematical, Physical and Engineering Sciences}\ }\textbf {\bibinfo {volume} {375}},\ \bibinfo {pages} {20160271} (\bibinfo {year} {2017})}\BibitemShut {NoStop}%
\bibitem [{\citenamefont {Yamada}\ \emph {et~al.}(2024)\citenamefont {Yamada}, \citenamefont {Nagakura}, \citenamefont {Akaho}, \citenamefont {Harada}, \citenamefont {Furusawa}, \citenamefont {Iwakami}, \citenamefont {Okawa}, \citenamefont {Matsufuru},\ and\ \citenamefont {Sumiyoshi}}]{yamada2024physical}%
  \BibitemOpen
  \bibfield  {author} {\bibinfo {author} {\bibfnamefont {S.}~\bibnamefont {Yamada}}, \bibinfo {author} {\bibfnamefont {H.}~\bibnamefont {Nagakura}}, \bibinfo {author} {\bibfnamefont {R.}~\bibnamefont {Akaho}}, \bibinfo {author} {\bibfnamefont {A.}~\bibnamefont {Harada}}, \bibinfo {author} {\bibfnamefont {S.}~\bibnamefont {Furusawa}}, \bibinfo {author} {\bibfnamefont {W.}~\bibnamefont {Iwakami}}, \bibinfo {author} {\bibfnamefont {H.}~\bibnamefont {Okawa}}, \bibinfo {author} {\bibfnamefont {H.}~\bibnamefont {Matsufuru}}, \ and\ \bibinfo {author} {\bibfnamefont {K.}~\bibnamefont {Sumiyoshi}},\ }\href@noop {} {\bibfield  {journal} {\bibinfo  {journal} {Proceedings of the Japan Academy, Series B}\ }\textbf {\bibinfo {volume} {100}},\ \bibinfo {pages} {190} (\bibinfo {year} {2024})}\BibitemShut {NoStop}%
\bibitem [{\citenamefont {Soker}(2022)}]{Soker:2022qxu}%
  \BibitemOpen
  \bibfield  {author} {\bibinfo {author} {\bibfnamefont {N.}~\bibnamefont {Soker}},\ }\href {\doibase 10.1088/1674-4527/ac7cbc} {\bibfield  {journal} {\bibinfo  {journal} {Res. Astron. Astrophys.}\ }\textbf {\bibinfo {volume} {22}},\ \bibinfo {pages} {095007} (\bibinfo {year} {2022})},\ \Eprint {http://arxiv.org/abs/2202.05556} {arXiv:2202.05556 [astro-ph.HE]} \BibitemShut {NoStop}%
\bibitem [{\citenamefont {Kuroda}\ and\ \citenamefont {Shibata}(2023)}]{Kuroda:2023zbz}%
  \BibitemOpen
  \bibfield  {author} {\bibinfo {author} {\bibfnamefont {T.}~\bibnamefont {Kuroda}}\ and\ \bibinfo {author} {\bibfnamefont {M.}~\bibnamefont {Shibata}},\ }\href {\doibase 10.1103/PhysRevD.107.103025} {\bibfield  {journal} {\bibinfo  {journal} {Phys. Rev. D}\ }\textbf {\bibinfo {volume} {107}},\ \bibinfo {pages} {103025} (\bibinfo {year} {2023})},\ \Eprint {http://arxiv.org/abs/2302.09853} {arXiv:2302.09853 [astro-ph.HE]} \BibitemShut {NoStop}%
\bibitem [{\citenamefont {Dasgupta}\ \emph {et~al.}(2012)\citenamefont {Dasgupta}, \citenamefont {O'Connor},\ and\ \citenamefont {Ott}}]{Dasgupta:2011jf}%
  \BibitemOpen
  \bibfield  {author} {\bibinfo {author} {\bibfnamefont {B.}~\bibnamefont {Dasgupta}}, \bibinfo {author} {\bibfnamefont {E.~P.}\ \bibnamefont {O'Connor}}, \ and\ \bibinfo {author} {\bibfnamefont {C.~D.}\ \bibnamefont {Ott}},\ }\href {\doibase 10.1103/PhysRevD.85.065008} {\bibfield  {journal} {\bibinfo  {journal} {Phys. Rev. D}\ }\textbf {\bibinfo {volume} {85}},\ \bibinfo {pages} {065008} (\bibinfo {year} {2012})},\ \Eprint {http://arxiv.org/abs/1106.1167} {arXiv:1106.1167 [astro-ph.SR]} \BibitemShut {NoStop}%
\bibitem [{\citenamefont {Pantaleone}(1992{\natexlab{a}})}]{PANTALEONE1992128}%
  \BibitemOpen
  \bibfield  {author} {\bibinfo {author} {\bibfnamefont {J.}~\bibnamefont {Pantaleone}},\ }\href {\doibase https://doi.org/10.1016/0370-2693(92)91887-F} {\bibfield  {journal} {\bibinfo  {journal} {Physics Letters B}\ }\textbf {\bibinfo {volume} {287}},\ \bibinfo {pages} {128} (\bibinfo {year} {1992}{\natexlab{a}})}\BibitemShut {NoStop}%
\bibitem [{\citenamefont {Pantaleone}(1992{\natexlab{b}})}]{Pantaleone:prd}%
  \BibitemOpen
  \bibfield  {author} {\bibinfo {author} {\bibfnamefont {J.}~\bibnamefont {Pantaleone}},\ }\href {\doibase 10.1103/PhysRevD.46.510} {\bibfield  {journal} {\bibinfo  {journal} {Phys. Rev. D}\ }\textbf {\bibinfo {volume} {46}},\ \bibinfo {pages} {510} (\bibinfo {year} {1992}{\natexlab{b}})}\BibitemShut {NoStop}%
\bibitem [{\citenamefont {Kostelecký}\ and\ \citenamefont {Samuel}(1993)}]{KOSTELECKY1993127}%
  \BibitemOpen
  \bibfield  {author} {\bibinfo {author} {\bibfnamefont {V.}~\bibnamefont {Kostelecký}}\ and\ \bibinfo {author} {\bibfnamefont {S.}~\bibnamefont {Samuel}},\ }\href {\doibase https://doi.org/10.1016/0370-2693(93)91795-O} {\bibfield  {journal} {\bibinfo  {journal} {Physics Letters B}\ }\textbf {\bibinfo {volume} {318}},\ \bibinfo {pages} {127} (\bibinfo {year} {1993})}\BibitemShut {NoStop}%
\bibitem [{\citenamefont {Samuel}(1993)}]{Samuel:1993uw}%
  \BibitemOpen
  \bibfield  {author} {\bibinfo {author} {\bibfnamefont {S.}~\bibnamefont {Samuel}},\ }\href {\doibase 10.1103/PhysRevD.48.1462} {\bibfield  {journal} {\bibinfo  {journal} {Phys. Rev. D}\ }\textbf {\bibinfo {volume} {48}},\ \bibinfo {pages} {1462} (\bibinfo {year} {1993})}\BibitemShut {NoStop}%
\bibitem [{\citenamefont {Kosteleck\'y}\ and\ \citenamefont {Samuel}(1995)}]{PhysRevD.52.621}%
  \BibitemOpen
  \bibfield  {author} {\bibinfo {author} {\bibfnamefont {V.~A.}\ \bibnamefont {Kosteleck\'y}}\ and\ \bibinfo {author} {\bibfnamefont {S.}~\bibnamefont {Samuel}},\ }\href {\doibase 10.1103/PhysRevD.52.621} {\bibfield  {journal} {\bibinfo  {journal} {Phys. Rev. D}\ }\textbf {\bibinfo {volume} {52}},\ \bibinfo {pages} {621} (\bibinfo {year} {1995})}\BibitemShut {NoStop}%
\bibitem [{\citenamefont {Qian}\ and\ \citenamefont {Fuller}(1995)}]{Qian:1994wh}%
  \BibitemOpen
  \bibfield  {author} {\bibinfo {author} {\bibfnamefont {Y.~Z.}\ \bibnamefont {Qian}}\ and\ \bibinfo {author} {\bibfnamefont {G.~M.}\ \bibnamefont {Fuller}},\ }\href {\doibase 10.1103/PhysRevD.51.1479} {\bibfield  {journal} {\bibinfo  {journal} {Phys. Rev. D}\ }\textbf {\bibinfo {volume} {51}},\ \bibinfo {pages} {1479} (\bibinfo {year} {1995})},\ \Eprint {http://arxiv.org/abs/astro-ph/9406073} {arXiv:astro-ph/9406073} \BibitemShut {NoStop}%
\bibitem [{\citenamefont {Pantaleone}(1995)}]{Pantaleone:1994ns}%
  \BibitemOpen
  \bibfield  {author} {\bibinfo {author} {\bibfnamefont {J.~T.}\ \bibnamefont {Pantaleone}},\ }\href {\doibase 10.1016/0370-2693(94)01369-N} {\bibfield  {journal} {\bibinfo  {journal} {Phys. Lett. B}\ }\textbf {\bibinfo {volume} {342}},\ \bibinfo {pages} {250} (\bibinfo {year} {1995})},\ \Eprint {http://arxiv.org/abs/astro-ph/9405008} {arXiv:astro-ph/9405008} \BibitemShut {NoStop}%
\bibitem [{\citenamefont {Kostelecky}\ and\ \citenamefont {Samuel}(1995)}]{Kostelecky:1994dt}%
  \BibitemOpen
  \bibfield  {author} {\bibinfo {author} {\bibfnamefont {V.~A.}\ \bibnamefont {Kostelecky}}\ and\ \bibinfo {author} {\bibfnamefont {S.}~\bibnamefont {Samuel}},\ }\href {\doibase 10.1103/PhysRevD.52.621} {\bibfield  {journal} {\bibinfo  {journal} {Phys. Rev. D}\ }\textbf {\bibinfo {volume} {52}},\ \bibinfo {pages} {621} (\bibinfo {year} {1995})},\ \Eprint {http://arxiv.org/abs/hep-ph/9506262} {arXiv:hep-ph/9506262} \BibitemShut {NoStop}%
\bibitem [{\citenamefont {Duan}\ \emph {et~al.}(2006{\natexlab{a}})\citenamefont {Duan}, \citenamefont {Fuller},\ and\ \citenamefont {Qian}}]{PhysRevD.74.123004}%
  \BibitemOpen
  \bibfield  {author} {\bibinfo {author} {\bibfnamefont {H.}~\bibnamefont {Duan}}, \bibinfo {author} {\bibfnamefont {G.~M.}\ \bibnamefont {Fuller}}, \ and\ \bibinfo {author} {\bibfnamefont {Y.-Z.}\ \bibnamefont {Qian}},\ }\href {\doibase 10.1103/PhysRevD.74.123004} {\bibfield  {journal} {\bibinfo  {journal} {Phys. Rev. D}\ }\textbf {\bibinfo {volume} {74}},\ \bibinfo {pages} {123004} (\bibinfo {year} {2006}{\natexlab{a}})}\BibitemShut {NoStop}%
\bibitem [{\citenamefont {Hannestad}\ \emph {et~al.}(2006)\citenamefont {Hannestad}, \citenamefont {Raffelt}, \citenamefont {Sigl},\ and\ \citenamefont {Wong}}]{PhysRevD.74.105010}%
  \BibitemOpen
  \bibfield  {author} {\bibinfo {author} {\bibfnamefont {S.}~\bibnamefont {Hannestad}}, \bibinfo {author} {\bibfnamefont {G.~G.}\ \bibnamefont {Raffelt}}, \bibinfo {author} {\bibfnamefont {G.}~\bibnamefont {Sigl}}, \ and\ \bibinfo {author} {\bibfnamefont {Y.~Y.~Y.}\ \bibnamefont {Wong}},\ }\href {\doibase 10.1103/PhysRevD.74.105010} {\bibfield  {journal} {\bibinfo  {journal} {Phys. Rev. D}\ }\textbf {\bibinfo {volume} {74}},\ \bibinfo {pages} {105010} (\bibinfo {year} {2006})}\BibitemShut {NoStop}%
\bibitem [{\citenamefont {Pastor}\ \emph {et~al.}(2002)\citenamefont {Pastor}, \citenamefont {Raffelt},\ and\ \citenamefont {Semikoz}}]{Pastor:2001iu}%
  \BibitemOpen
  \bibfield  {author} {\bibinfo {author} {\bibfnamefont {S.}~\bibnamefont {Pastor}}, \bibinfo {author} {\bibfnamefont {G.~G.}\ \bibnamefont {Raffelt}}, \ and\ \bibinfo {author} {\bibfnamefont {D.~V.}\ \bibnamefont {Semikoz}},\ }\href {\doibase 10.1103/PhysRevD.65.053011} {\bibfield  {journal} {\bibinfo  {journal} {Phys. Rev. D}\ }\textbf {\bibinfo {volume} {65}},\ \bibinfo {pages} {053011} (\bibinfo {year} {2002})},\ \Eprint {http://arxiv.org/abs/hep-ph/0109035} {arXiv:hep-ph/0109035} \BibitemShut {NoStop}%
\bibitem [{\citenamefont {Duan}\ \emph {et~al.}(2010)\citenamefont {Duan}, \citenamefont {Fuller},\ and\ \citenamefont {Qian}}]{Duan:2010bg}%
  \BibitemOpen
  \bibfield  {author} {\bibinfo {author} {\bibfnamefont {H.}~\bibnamefont {Duan}}, \bibinfo {author} {\bibfnamefont {G.~M.}\ \bibnamefont {Fuller}}, \ and\ \bibinfo {author} {\bibfnamefont {Y.-Z.}\ \bibnamefont {Qian}},\ }\href {\doibase 10.1146/annurev.nucl.012809.104524} {\bibfield  {journal} {\bibinfo  {journal} {Ann. Rev. Nucl. Part. Sci.}\ }\textbf {\bibinfo {volume} {60}},\ \bibinfo {pages} {569} (\bibinfo {year} {2010})},\ \Eprint {http://arxiv.org/abs/1001.2799} {arXiv:1001.2799 [hep-ph]} \BibitemShut {NoStop}%
\bibitem [{\citenamefont {Pehlivan}\ \emph {et~al.}(2011)\citenamefont {Pehlivan}, \citenamefont {Balantekin}, \citenamefont {Kajino},\ and\ \citenamefont {Yoshida}}]{Pehlivan:2011hp}%
  \BibitemOpen
  \bibfield  {author} {\bibinfo {author} {\bibfnamefont {Y.}~\bibnamefont {Pehlivan}}, \bibinfo {author} {\bibfnamefont {A.~B.}\ \bibnamefont {Balantekin}}, \bibinfo {author} {\bibfnamefont {T.}~\bibnamefont {Kajino}}, \ and\ \bibinfo {author} {\bibfnamefont {T.}~\bibnamefont {Yoshida}},\ }\href {\doibase 10.1103/PhysRevD.84.065008} {\bibfield  {journal} {\bibinfo  {journal} {Phys. Rev. D}\ }\textbf {\bibinfo {volume} {84}},\ \bibinfo {pages} {065008} (\bibinfo {year} {2011})},\ \Eprint {http://arxiv.org/abs/1105.1182} {arXiv:1105.1182 [astro-ph.CO]} \BibitemShut {NoStop}%
\bibitem [{\citenamefont {Patwardhan}\ \emph {et~al.}(2023)\citenamefont {Patwardhan}, \citenamefont {Cervia}, \citenamefont {Rrapaj}, \citenamefont {Siwach},\ and\ \citenamefont {Balantekin}}]{Patwardhan:2022mxg}%
  \BibitemOpen
  \bibfield  {author} {\bibinfo {author} {\bibfnamefont {A.~V.}\ \bibnamefont {Patwardhan}}, \bibinfo {author} {\bibfnamefont {M.~J.}\ \bibnamefont {Cervia}}, \bibinfo {author} {\bibfnamefont {E.}~\bibnamefont {Rrapaj}}, \bibinfo {author} {\bibfnamefont {P.}~\bibnamefont {Siwach}}, \ and\ \bibinfo {author} {\bibfnamefont {A.~B.}\ \bibnamefont {Balantekin}},\ }\enquote {\bibinfo {title} {{Many-Body Collective Neutrino Oscillations: Recent Developments}},}\ in\ \href {\doibase 10.1007/978-981-15-8818-1_126-1} {\emph {\bibinfo {booktitle} {{Handbook of Nuclear Physics}}}},\ \bibinfo {editor} {edited by\ \bibinfo {editor} {\bibfnamefont {I.}~\bibnamefont {Tanihata}}, \bibinfo {editor} {\bibfnamefont {H.}~\bibnamefont {Toki}}, \ and\ \bibinfo {editor} {\bibfnamefont {T.}~\bibnamefont {Kajino}}}\ (\bibinfo  {publisher} {Springer},\ \bibinfo {year} {2023})\ pp.\ \bibinfo {pages} {1--16},\ \Eprint {http://arxiv.org/abs/2301.00342} {arXiv:2301.00342 [hep-ph]} \BibitemShut {NoStop}%
\bibitem [{\citenamefont {Hall}\ \emph {et~al.}(2021)\citenamefont {Hall}, \citenamefont {Roggero}, \citenamefont {Baroni},\ and\ \citenamefont {Carlson}}]{Hall:2021rbv}%
  \BibitemOpen
  \bibfield  {author} {\bibinfo {author} {\bibfnamefont {B.}~\bibnamefont {Hall}}, \bibinfo {author} {\bibfnamefont {A.}~\bibnamefont {Roggero}}, \bibinfo {author} {\bibfnamefont {A.}~\bibnamefont {Baroni}}, \ and\ \bibinfo {author} {\bibfnamefont {J.}~\bibnamefont {Carlson}},\ }\href {\doibase 10.1103/PhysRevD.104.063009} {\bibfield  {journal} {\bibinfo  {journal} {Phys. Rev. D}\ }\textbf {\bibinfo {volume} {104}},\ \bibinfo {pages} {063009} (\bibinfo {year} {2021})},\ \Eprint {http://arxiv.org/abs/2102.12556} {arXiv:2102.12556 [quant-ph]} \BibitemShut {NoStop}%
\bibitem [{\citenamefont {Delfan~Azari}\ \emph {et~al.}(2020)\citenamefont {Delfan~Azari}, \citenamefont {Yamada}, \citenamefont {Morinaga}, \citenamefont {Nagakura}, \citenamefont {Furusawa}, \citenamefont {Harada}, \citenamefont {Okawa}, \citenamefont {Iwakami},\ and\ \citenamefont {Sumiyoshi}}]{DelfanAzari:2019tez}%
  \BibitemOpen
  \bibfield  {author} {\bibinfo {author} {\bibfnamefont {M.}~\bibnamefont {Delfan~Azari}}, \bibinfo {author} {\bibfnamefont {S.}~\bibnamefont {Yamada}}, \bibinfo {author} {\bibfnamefont {T.}~\bibnamefont {Morinaga}}, \bibinfo {author} {\bibfnamefont {H.}~\bibnamefont {Nagakura}}, \bibinfo {author} {\bibfnamefont {S.}~\bibnamefont {Furusawa}}, \bibinfo {author} {\bibfnamefont {A.}~\bibnamefont {Harada}}, \bibinfo {author} {\bibfnamefont {H.}~\bibnamefont {Okawa}}, \bibinfo {author} {\bibfnamefont {W.}~\bibnamefont {Iwakami}}, \ and\ \bibinfo {author} {\bibfnamefont {K.}~\bibnamefont {Sumiyoshi}},\ }\href {\doibase 10.1103/PhysRevD.101.023018} {\bibfield  {journal} {\bibinfo  {journal} {Phys. Rev. D}\ }\textbf {\bibinfo {volume} {101}},\ \bibinfo {pages} {023018} (\bibinfo {year} {2020})},\ \Eprint {http://arxiv.org/abs/1910.06176} {arXiv:1910.06176 [astro-ph.HE]} \BibitemShut {NoStop}%
\bibitem [{\citenamefont {Gava}\ and\ \citenamefont {Volpe}(2008)}]{Gava:2008rp}%
  \BibitemOpen
  \bibfield  {author} {\bibinfo {author} {\bibfnamefont {J.}~\bibnamefont {Gava}}\ and\ \bibinfo {author} {\bibfnamefont {C.}~\bibnamefont {Volpe}},\ }\href {\doibase 10.1103/PhysRevD.78.083007} {\bibfield  {journal} {\bibinfo  {journal} {Phys. Rev. D}\ }\textbf {\bibinfo {volume} {78}},\ \bibinfo {pages} {083007} (\bibinfo {year} {2008})},\ \Eprint {http://arxiv.org/abs/0807.3418} {arXiv:0807.3418 [astro-ph]} \BibitemShut {NoStop}%
\bibitem [{\citenamefont {Duan}\ and\ \citenamefont {Friedland}(2011)}]{Duan:2010bf}%
  \BibitemOpen
  \bibfield  {author} {\bibinfo {author} {\bibfnamefont {H.}~\bibnamefont {Duan}}\ and\ \bibinfo {author} {\bibfnamefont {A.}~\bibnamefont {Friedland}},\ }\href {\doibase 10.1103/PhysRevLett.106.091101} {\bibfield  {journal} {\bibinfo  {journal} {Phys. Rev. Lett.}\ }\textbf {\bibinfo {volume} {106}},\ \bibinfo {pages} {091101} (\bibinfo {year} {2011})},\ \Eprint {http://arxiv.org/abs/1006.2359} {arXiv:1006.2359 [hep-ph]} \BibitemShut {NoStop}%
\bibitem [{\citenamefont {Chatelain}\ and\ \citenamefont {Volpe}(2020)}]{Chatelain:2019nkf}%
  \BibitemOpen
  \bibfield  {author} {\bibinfo {author} {\bibfnamefont {A.}~\bibnamefont {Chatelain}}\ and\ \bibinfo {author} {\bibfnamefont {M.~C.}\ \bibnamefont {Volpe}},\ }\href {\doibase 10.1016/j.physletb.2019.135150} {\bibfield  {journal} {\bibinfo  {journal} {Phys. Lett. B}\ }\textbf {\bibinfo {volume} {801}},\ \bibinfo {pages} {135150} (\bibinfo {year} {2020})},\ \Eprint {http://arxiv.org/abs/1906.12152} {arXiv:1906.12152 [hep-ph]} \BibitemShut {NoStop}%
\bibitem [{\citenamefont {Chakrabarty}\ and\ \citenamefont {Lahiri}(2019)}]{Chakrabarty:2019cau}%
  \BibitemOpen
  \bibfield  {author} {\bibinfo {author} {\bibfnamefont {S.}~\bibnamefont {Chakrabarty}}\ and\ \bibinfo {author} {\bibfnamefont {A.}~\bibnamefont {Lahiri}},\ }\href {\doibase 10.1140/epjc/s10052-019-7209-2} {\bibfield  {journal} {\bibinfo  {journal} {Eur. Phys. J. C}\ }\textbf {\bibinfo {volume} {79}},\ \bibinfo {pages} {697} (\bibinfo {year} {2019})},\ \Eprint {http://arxiv.org/abs/1904.06036} {arXiv:1904.06036 [hep-ph]} \BibitemShut {NoStop}%
\bibitem [{\citenamefont {Barick}\ \emph {et~al.}(2023)\citenamefont {Barick}, \citenamefont {Ghose},\ and\ \citenamefont {Lahiri}}]{Barick:2023qjq}%
  \BibitemOpen
  \bibfield  {author} {\bibinfo {author} {\bibfnamefont {R.}~\bibnamefont {Barick}}, \bibinfo {author} {\bibfnamefont {I.}~\bibnamefont {Ghose}}, \ and\ \bibinfo {author} {\bibfnamefont {A.}~\bibnamefont {Lahiri}},\ }\href {\doibase 10.31526/lhep.2023.362} {\bibfield  {journal} {\bibinfo  {journal} {LHEP}\ }\textbf {\bibinfo {volume} {2023}},\ \bibinfo {pages} {362} (\bibinfo {year} {2023})},\ \Eprint {http://arxiv.org/abs/2305.05903} {arXiv:2305.05903 [hep-ph]} \BibitemShut {NoStop}%
\bibitem [{\citenamefont {Barick}\ \emph {et~al.}(2024)\citenamefont {Barick}, \citenamefont {Ghose},\ and\ \citenamefont {Lahiri}}]{Barick:2023wxx}%
  \BibitemOpen
  \bibfield  {author} {\bibinfo {author} {\bibfnamefont {R.}~\bibnamefont {Barick}}, \bibinfo {author} {\bibfnamefont {I.}~\bibnamefont {Ghose}}, \ and\ \bibinfo {author} {\bibfnamefont {A.}~\bibnamefont {Lahiri}},\ }\href {\doibase 10.1140/epjp/s13360-024-05296-8} {\bibfield  {journal} {\bibinfo  {journal} {Eur. Phys. J. Plus}\ }\textbf {\bibinfo {volume} {139}},\ \bibinfo {pages} {461} (\bibinfo {year} {2024})},\ \Eprint {http://arxiv.org/abs/2302.10945} {arXiv:2302.10945 [hep-ph]} \BibitemShut {NoStop}%
\bibitem [{\citenamefont {Panda}\ \emph {et~al.}(2025)\citenamefont {Panda}, \citenamefont {Singha}, \citenamefont {Ghosh},\ and\ \citenamefont {Mohanta}}]{Panda:2024qsh}%
  \BibitemOpen
  \bibfield  {author} {\bibinfo {author} {\bibfnamefont {P.}~\bibnamefont {Panda}}, \bibinfo {author} {\bibfnamefont {D.~K.}\ \bibnamefont {Singha}}, \bibinfo {author} {\bibfnamefont {M.}~\bibnamefont {Ghosh}}, \ and\ \bibinfo {author} {\bibfnamefont {R.}~\bibnamefont {Mohanta}},\ }\href {\doibase 10.1140/epjc/s10052-025-13771-4} {\bibfield  {journal} {\bibinfo  {journal} {Eur. Phys. J. C}\ }\textbf {\bibinfo {volume} {85}},\ \bibinfo {pages} {67} (\bibinfo {year} {2025})},\ \Eprint {http://arxiv.org/abs/2403.09105} {arXiv:2403.09105 [hep-ph]} \BibitemShut {NoStop}%
\bibitem [{\citenamefont {Cartan}(1923)}]{Cartan:1923zea}%
  \BibitemOpen
  \bibfield  {author} {\bibinfo {author} {\bibfnamefont {E.}~\bibnamefont {Cartan}},\ }\href@noop {} {\bibfield  {journal} {\bibinfo  {journal} {Annales Sci. Ecole Norm. Sup.}\ }\textbf {\bibinfo {volume} {40}},\ \bibinfo {pages} {325} (\bibinfo {year} {1923})}\BibitemShut {NoStop}%
\bibitem [{\citenamefont {Hehl}\ \emph {et~al.}(1976)\citenamefont {Hehl}, \citenamefont {Von Der~Heyde}, \citenamefont {Kerlick},\ and\ \citenamefont {Nester}}]{Hehl:1976kj}%
  \BibitemOpen
  \bibfield  {author} {\bibinfo {author} {\bibfnamefont {F.~W.}\ \bibnamefont {Hehl}}, \bibinfo {author} {\bibfnamefont {P.}~\bibnamefont {Von Der~Heyde}}, \bibinfo {author} {\bibfnamefont {G.~D.}\ \bibnamefont {Kerlick}}, \ and\ \bibinfo {author} {\bibfnamefont {J.~M.}\ \bibnamefont {Nester}},\ }\href {\doibase 10.1103/RevModPhys.48.393} {\bibfield  {journal} {\bibinfo  {journal} {Rev. Mod. Phys.}\ }\textbf {\bibinfo {volume} {48}},\ \bibinfo {pages} {393} (\bibinfo {year} {1976})}\BibitemShut {NoStop}%
\bibitem [{\citenamefont {Carroll}(2019)}]{Carroll:2004st}%
  \BibitemOpen
  \bibfield  {author} {\bibinfo {author} {\bibfnamefont {S.~M.}\ \bibnamefont {Carroll}},\ }\href@noop {} {\emph {\bibinfo {title} {{Spacetime and Geometry}: {An Introduction to General Relativity}}}}\ (\bibinfo  {publisher} {Cambridge University Press},\ \bibinfo {year} {2019})\BibitemShut {NoStop}%
\bibitem [{\citenamefont {Workman}\ \emph {et~al.}(2022)\citenamefont {Workman} \emph {et~al.}}]{ParticleDataGroup:2022pth}%
  \BibitemOpen
  \bibfield  {author} {\bibinfo {author} {\bibfnamefont {R.~L.}\ \bibnamefont {Workman}} \emph {et~al.} (\bibinfo {collaboration} {Particle Data Group}),\ }\href {\doibase 10.1093/ptep/ptac097} {\bibfield  {journal} {\bibinfo  {journal} {PTEP}\ }\textbf {\bibinfo {volume} {2022}},\ \bibinfo {pages} {083C01} (\bibinfo {year} {2022})}\BibitemShut {NoStop}%
\bibitem [{\citenamefont {Wolfenstein}(1978)}]{PhysRevD.17.2369}%
  \BibitemOpen
  \bibfield  {author} {\bibinfo {author} {\bibfnamefont {L.}~\bibnamefont {Wolfenstein}},\ }\href {\doibase 10.1103/PhysRevD.17.2369} {\bibfield  {journal} {\bibinfo  {journal} {Phys. Rev. D}\ }\textbf {\bibinfo {volume} {17}},\ \bibinfo {pages} {2369} (\bibinfo {year} {1978})}\BibitemShut {NoStop}%
\bibitem [{\citenamefont {Ghose}\ \emph {et~al.}(2023)\citenamefont {Ghose}, \citenamefont {Barick},\ and\ \citenamefont {Lahiri}}]{Ghose:2023ttq}%
  \BibitemOpen
  \bibfield  {author} {\bibinfo {author} {\bibfnamefont {I.}~\bibnamefont {Ghose}}, \bibinfo {author} {\bibfnamefont {R.}~\bibnamefont {Barick}}, \ and\ \bibinfo {author} {\bibfnamefont {A.}~\bibnamefont {Lahiri}},\ }\href {\doibase 10.31526/lhep.2023.349} {\bibfield  {journal} {\bibinfo  {journal} {LHEP}\ }\textbf {\bibinfo {volume} {2023}},\ \bibinfo {pages} {349} (\bibinfo {year} {2023})},\ \Eprint {http://arxiv.org/abs/2302.10119} {arXiv:2302.10119 [hep-ph]} \BibitemShut {NoStop}%
\bibitem [{\citenamefont {Pal}\ and\ \citenamefont {Pham}(1989)}]{Pal:1989xs}%
  \BibitemOpen
  \bibfield  {author} {\bibinfo {author} {\bibfnamefont {P.~B.}\ \bibnamefont {Pal}}\ and\ \bibinfo {author} {\bibfnamefont {T.~N.}\ \bibnamefont {Pham}},\ }\href {\doibase 10.1103/PhysRevD.40.259} {\bibfield  {journal} {\bibinfo  {journal} {Phys. Rev. D}\ }\textbf {\bibinfo {volume} {40}},\ \bibinfo {pages} {259} (\bibinfo {year} {1989})}\BibitemShut {NoStop}%
\bibitem [{\citenamefont {Lin}\ and\ \citenamefont {Duan}(2023)}]{Lin:2022dek}%
  \BibitemOpen
  \bibfield  {author} {\bibinfo {author} {\bibfnamefont {Y.-C.}\ \bibnamefont {Lin}}\ and\ \bibinfo {author} {\bibfnamefont {H.}~\bibnamefont {Duan}},\ }\href {\doibase 10.1103/PhysRevD.107.083034} {\bibfield  {journal} {\bibinfo  {journal} {Phys. Rev. D}\ }\textbf {\bibinfo {volume} {107}},\ \bibinfo {pages} {083034} (\bibinfo {year} {2023})},\ \Eprint {http://arxiv.org/abs/2210.09218} {arXiv:2210.09218 [hep-ph]} \BibitemShut {NoStop}%
\bibitem [{\citenamefont {Cyr-Racine}\ and\ \citenamefont {Sigurdson}(2014)}]{Cyr-Racine:2013jua}%
  \BibitemOpen
  \bibfield  {author} {\bibinfo {author} {\bibfnamefont {F.-Y.}\ \bibnamefont {Cyr-Racine}}\ and\ \bibinfo {author} {\bibfnamefont {K.}~\bibnamefont {Sigurdson}},\ }\href {\doibase 10.1103/PhysRevD.90.123533} {\bibfield  {journal} {\bibinfo  {journal} {Phys. Rev. D}\ }\textbf {\bibinfo {volume} {90}},\ \bibinfo {pages} {123533} (\bibinfo {year} {2014})},\ \Eprint {http://arxiv.org/abs/1306.1536} {arXiv:1306.1536 [astro-ph.CO]} \BibitemShut {NoStop}%
\bibitem [{\citenamefont {Camarena}\ \emph {et~al.}(2023)\citenamefont {Camarena}, \citenamefont {Cyr-Racine},\ and\ \citenamefont {Houghteling}}]{Camarena:2023cku}%
  \BibitemOpen
  \bibfield  {author} {\bibinfo {author} {\bibfnamefont {D.}~\bibnamefont {Camarena}}, \bibinfo {author} {\bibfnamefont {F.-Y.}\ \bibnamefont {Cyr-Racine}}, \ and\ \bibinfo {author} {\bibfnamefont {J.}~\bibnamefont {Houghteling}},\ }\href {\doibase 10.1103/PhysRevD.108.103535} {\bibfield  {journal} {\bibinfo  {journal} {Phys. Rev. D}\ }\textbf {\bibinfo {volume} {108}},\ \bibinfo {pages} {103535} (\bibinfo {year} {2023})},\ \Eprint {http://arxiv.org/abs/2309.03941} {arXiv:2309.03941 [astro-ph.CO]} \BibitemShut {NoStop}%
\bibitem [{\citenamefont {Das}\ and\ \citenamefont {Ghosh}(2023)}]{Das:2023npl}%
  \BibitemOpen
  \bibfield  {author} {\bibinfo {author} {\bibfnamefont {A.}~\bibnamefont {Das}}\ and\ \bibinfo {author} {\bibfnamefont {S.}~\bibnamefont {Ghosh}},\ }\href {\doibase 10.1088/1475-7516/2023/09/042} {\bibfield  {journal} {\bibinfo  {journal} {JCAP}\ }\textbf {\bibinfo {volume} {09}},\ \bibinfo {pages} {042} (\bibinfo {year} {2023})},\ \Eprint {http://arxiv.org/abs/2303.08843} {arXiv:2303.08843 [astro-ph.CO]} \BibitemShut {NoStop}%
\bibitem [{\citenamefont {Duan}\ \emph {et~al.}(2006{\natexlab{b}})\citenamefont {Duan}, \citenamefont {Fuller}, \citenamefont {Carlson},\ and\ \citenamefont {Qian}}]{Duan:2006an}%
  \BibitemOpen
  \bibfield  {author} {\bibinfo {author} {\bibfnamefont {H.}~\bibnamefont {Duan}}, \bibinfo {author} {\bibfnamefont {G.~M.}\ \bibnamefont {Fuller}}, \bibinfo {author} {\bibfnamefont {J.}~\bibnamefont {Carlson}}, \ and\ \bibinfo {author} {\bibfnamefont {Y.-Z.}\ \bibnamefont {Qian}},\ }\href {\doibase 10.1103/PhysRevD.74.105014} {\bibfield  {journal} {\bibinfo  {journal} {Phys. Rev. D}\ }\textbf {\bibinfo {volume} {74}},\ \bibinfo {pages} {105014} (\bibinfo {year} {2006}{\natexlab{b}})},\ \Eprint {http://arxiv.org/abs/astro-ph/0606616} {arXiv:astro-ph/0606616} \BibitemShut {NoStop}%
\bibitem [{\citenamefont {Nieves}\ and\ \citenamefont {Pal}(2004)}]{Nieves:2004gfi}%
  \BibitemOpen
  \bibfield  {author} {\bibinfo {author} {\bibfnamefont {J.~F.}\ \bibnamefont {Nieves}}\ and\ \bibinfo {author} {\bibfnamefont {P.~B.}\ \bibnamefont {Pal}},\ }\href {\doibase 10.1119/1.1757445} {\bibfield  {journal} {\bibinfo  {journal} {American Journal of Physics}\ }\textbf {\bibinfo {volume} {72}},\ \bibinfo {pages} {1100} (\bibinfo {year} {2004})},\ \Eprint {http://arxiv.org/abs/https://doi.org/10.1119/1.1757445} {https://doi.org/10.1119/1.1757445} \BibitemShut {NoStop}%
\bibitem [{\citenamefont {Dasgupta}\ and\ \citenamefont {Dighe}(2008)}]{Dasgupta:2007ws}%
  \BibitemOpen
  \bibfield  {author} {\bibinfo {author} {\bibfnamefont {B.}~\bibnamefont {Dasgupta}}\ and\ \bibinfo {author} {\bibfnamefont {A.}~\bibnamefont {Dighe}},\ }\href {\doibase 10.1103/PhysRevD.77.113002} {\bibfield  {journal} {\bibinfo  {journal} {Phys. Rev. D}\ }\textbf {\bibinfo {volume} {77}},\ \bibinfo {pages} {113002} (\bibinfo {year} {2008})},\ \Eprint {http://arxiv.org/abs/0712.3798} {arXiv:0712.3798 [hep-ph]} \BibitemShut {NoStop}%
\bibitem [{\citenamefont {Neto}\ and\ \citenamefont {Kemp}(2022)}]{Neto:2021hhl}%
  \BibitemOpen
  \bibfield  {author} {\bibinfo {author} {\bibfnamefont {P.~D.}\ \bibnamefont {Neto}}\ and\ \bibinfo {author} {\bibfnamefont {E.}~\bibnamefont {Kemp}},\ }\href {\doibase 10.1142/S0217732322500481} {\bibfield  {journal} {\bibinfo  {journal} {Mod. Phys. Lett. A}\ }\textbf {\bibinfo {volume} {37}},\ \bibinfo {pages} {2250048} (\bibinfo {year} {2022})},\ \Eprint {http://arxiv.org/abs/2111.11480} {arXiv:2111.11480 [hep-ph]} \BibitemShut {NoStop}%
\end{thebibliography}%
\bibliographystyle{apsrev4-1}

\end{document}